\documentclass{article}

\usepackage{microtype}
\usepackage{graphicx}
\usepackage{placeins}
\usepackage{needspace}
\usepackage{subcaption}
\usepackage{booktabs}

\newcommand{\maintablestyle}{\small\renewcommand{\arraystretch}{1.12}\setlength{\tabcolsep}{3pt}}
\newcommand{\appendixtablestyle}{\small\renewcommand{\arraystretch}{1.12}\setlength{\tabcolsep}{4pt}}

\usepackage{hyperref}

\usepackage[preprint]{icml2026}

\usepackage{amsmath}
\usepackage{amssymb}
\usepackage{mathtools}
\usepackage{amsthm}
\usepackage[capitalize,noabbrev,nameinlink]{cleveref}
\usepackage[textsize=tiny]{todonotes}

\icmltitlerunning{SimpFlow: Transferable Pretraining for Fluid Dynamics via Simplified Physics}

\begin{document}

\twocolumn[
  \icmltitle{SimpFlow: Transferable Pretraining for Fluid Dynamics via Simplified Physics}

  \begin{icmlauthorlist}
    \icmlauthor{Ruijie Li}{imech,ucas}
    \icmlauthor{Tianbai Xiao}{imech,ucas}
  \end{icmlauthorlist}
  \icmlaffiliation{imech}{Centre for Interdisciplinary Research in Fluids, Institute of Mechanics, Chinese Academy of Sciences, Beijing, China}
  \icmlaffiliation{ucas}{School of Engineering Science, University of Chinese Academy of Sciences, Beijing, China}
  \icmlcorrespondingauthor{Tianbai Xiao}{txiao@imech.ac.cn}
  \icmlkeywords{Neural Simulators, Pretraining, Computational Fluid Dynamics, Potential Flow}

  \vskip 0.3in
]

\printAffiliationsAndNotice{}
\suppressfloats[t]

\begin{abstract}
Neural surrogates enable rapid fluid simulation, but learning across diverse three-dimensional (3D) geometries requires costly computational fluid dynamics (CFD) labels.
We introduce SimpFlow, a pretraining approach that generates supervision from simplified physics on unlabeled geometry libraries.
A source-panel potential-flow solver generates velocity and pressure fields for approximately 9,000 3D geometries, capturing geometry-induced acceleration, stagnation, and suction while omitting viscosity and separated wakes.
The solver imports only a surface mesh and reuses precomputed velocity fields across inflow directions, enabling inexpensive label generation.
In our benchmark, amortized labeling takes 0.111 s per example using four CPU threads. 
We pretrain a neural surrogate on these labels and fine-tune it on independent CFD datasets.
Across five flow tasks, SimpFlow improves over training from scratch in all 24 task–budget settings. 
With fine-tuning fixed at 200 epochs, interpolation indicates that SimpFlow requires 35–68\% fewer CFD training cases to match the error of scratch models trained at each task’s largest data budget. 
These results demonstrate that simplified physical solutions can provide transferable supervision, 
%despite large pointwise label errors
offering a practical way to reduce the high-fidelity data requirements of fluid surrogates.
\end{abstract}
% Unnumbered first-page right-column summary; full Figure 3 is unchanged.
\begin{figure}[t]
\centering
\includegraphics[width=\columnwidth]{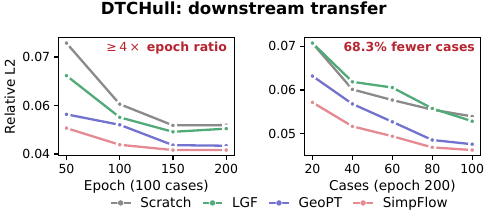}
\par{\small\bfseries Greater transfer gains\par}
\vspace{5pt}
{\small\bfseries Labeling time per example}\par\vspace{3pt}
{\maintablestyle
\begin{tabular}{llr}
\toprule
\textbf{Method} & \textbf{Hardware} & \textbf{s/example}\\
\midrule
SimpFlow (4k) & 4 CPU threads & \textbf{0.111}\\
\shortstack[l]{GeoPT\\\citep{wu2026geopt}} & 4 CPU threads & 0.309\\
\shortstack[l]{LBM\\\citep{shen2026scaling3d}} & 1 A100 GPU & 135.5$^{*}$\\
\bottomrule
\end{tabular}}\par\vspace{3pt}
{\small\bfseries Lower labeling cost\par}\vspace{3pt}
{\scriptsize\raggedright
CPU: AMD EPYC 9684X. $^{*}$LBM: reported solve time amortized over four query clouds ($542/4$).\par}
\vspace{5pt}
{\small\raggedright\textbf{64.2\% less labeling time in the benchmark than our GeoPT reproduction, with lower DTCHull error at every budget.}\par}
\vspace{2pt}
{\scriptsize\raggedright
Curves from \cref{fig:transfer}; savings and epoch ratios use the scratch reference accuracy (\cref{app:efficiency}).
Cost accounting and LBM source: \cref{tab:threewaycost}.\par}
\end{figure}

\section{Introduction}
Neural surrogate models provide an alternative for rapid fluid simulation~\citep{lu2021deeponet,li2021fno,li2023gino,wu2024transolver,alkin2024upt}.
After training, they can predict flow fields orders of magnitude faster than classical numerical solvers, making repeated evaluations practical for design iteration across 3D geometries.
Learning these models, however, requires costly CFD labels that cover diverse shapes and operating conditions. For example, a typical high-fidelity simulation requires approximately $6.1\times10^4$ CPU core-hours~\citep{ashton2024drivaerml}. Generating sufficient training data across a broad range of geometries therefore remains a substantial cost.

Pretraining learns reusable representations from source data and can reduce the high-fidelity data needed for adaptation to a target CFD task~\citep{bleeker2025neuralcfdv1,zhou2026crossdomain}. Its practical value also depends on the cost of generating the source supervision. Unlabeled geometry libraries provide broad shape coverage, but using them for flow pretraining requires supervision that is both inexpensive to generate and informative for downstream prediction.

Existing approaches provide several sources of supervision. Pretraining on PDE solution corpora, as in MPP, DPOT, and Poseidon, reuses simulated trajectories across problems~\citep{mccabe2024mpp,hao2024dpot,herde2024poseidon}; GP-UPT and cross-domain pretraining transfer representations learned from existing CFD datasets~\citep{bleeker2025neuralcfdv1,zhou2026crossdomain}. Geometry-based approaches use unlabeled shapes through auxiliary targets, including occupancy reconstruction and the synthetic geometric dynamics of GeoPT~\citep{zhang2025latentshape,wu2026geopt}. Multi-fidelity learning combines high-fidelity data with cheaper solutions from simplified equations or coarser discretizations~\citep{meng2020multifidelity,lu2022multifidelity,setinek2025multifidelity,setinek2026computeoptimal}. Within this direction, \citet{wong2025panel} demonstrate transfer from inviscid panel solutions to RANS predictions for a fixed NACA0012 airfoil, while concurrent work by \citet{shen2026scaling3d} pretrains on lattice-Boltzmann simulations of independent 3D source geometries. These studies motivate examining how inexpensive physical solves can support pretraining across diverse 3D geometries.

We introduce \emph{SimpFlow}, a pretraining approach that uses simplified physics to generate flow supervision from unlabeled geometry libraries. A source-panel potential-flow solver~\citep{hess1967panel} provides velocity and pressure-coefficient ($C_p$) labels on approximately 9,000 ShapeNet geometries spanning cars, airplanes, and watercraft~\citep{chang2015shapenet}. These solutions capture geometry-induced acceleration, stagnation, and suction while omitting viscosity and separated wakes. The solver requires only a surface mesh and reuses precomputed velocity fields across inflow directions, enabling inexpensive labeling across flow conditions. We pretrain Transolver~\citep{wu2024transolver} on these labels and fine-tune the resulting model on independent CFD datasets.

In our labeling benchmark, the 4k-panel solver takes an amortized 0.111\,s per example using four CPU threads (\cref{sec:labelcost}). Across five flow tasks, SimpFlow improves over training from scratch in all 24 task--budget settings. With fine-tuning fixed at 200 epochs, interpolation indicates that it requires 35--68\% fewer CFD training cases to match the error of scratch models trained at each task's largest data budget. Under the shared training protocol, it also lowers test error relative to GeoPT in 21 of the 24 settings (\cref{fig:transfer}).

To examine the role of supervision content, we compare potential-flow pretraining with a Local Geometry Field (LGF) control that assigns the same output quantities using local geometric rules. We also vary the panel count within the same potential-flow model to study how numerical resolution affects labeling cost and downstream transfer.

Our contributions are:
\begin{itemize}
  \item We introduce SimpFlow, a pretraining pipeline that labels diverse unlabeled 3D geometries with potential-flow velocity and pressure fields, using surface meshes and reusable inflow bases to reduce labeling work.
  \item We evaluate a shared initialization on five flow tasks built independently of the source geometry library, demonstrating lower test errors than scratch and lower interpolated CFD sample requirements under a fixed fine-tuning budget.
  \item We examine the roles of supervision content and numerical resolution through an LGF control and a panel-resolution study, characterizing their effects on transfer performance and labeling cost.
\end{itemize}

\section{Related Work}
\label{sec:related}

\subsection{Neural surrogates and physics-informed learning}
Neural operators learn mappings from input functions to PDE solutions, amortizing numerical simulation across a problem family. DeepONet~\citep{lu2021deeponet} combines input-function and query-coordinate representations, while FNO~\citep{li2021fno} learns integral operators in Fourier space. Attention-based approaches extend operator learning to heterogeneous inputs and spatial discretizations: GNOT~\citep{hao2023gnot} handles multiple input functions and irregular meshes, and Transolver~\citep{wu2024transolver} aggregates points into learned physical states for efficient interaction. UPT~\citep{alkin2024upt} processes simulations in a compact latent space, while AB-UPT~\citep{alkin2025abupt} scales surface and volume prediction using branched operators and anchored field decoders. We use Transolver to study how inexpensive physical supervision shapes the initialization for downstream flow prediction.

Physics-informed learning incorporates physical knowledge into model training, providing additional constraints beyond observed solution data~\citep{karniadakis2021piml}.
Physics-informed neural networks (PINNs) approximate PDE solutions with neural networks and construct losses from governing-equation residuals, prescribed initial and boundary conditions, and available observations~\citep{raissi2019pinn}.
Automatic differentiation supplies the derivatives needed to evaluate PDE residuals at unlabeled collocation points.
Physics-informed neural operator (PINO) extends this principle to operator learning by combining coarse-resolution solution data with PDE constraints evaluated at a finer resolution~\citep{li2024pino}.
It learns solution operators across a family of PDE instances and demonstrates prediction at resolutions beyond those of the training data.

\subsection{Geometry-based pretraining}
Geometry-based pretraining provides source data without physical solution labels. GeoPT~\citep{wu2026geopt} lifts 3D geometries with synthetic dynamics and learns geometry-bounded trajectories; its analysis also shows that static geometry-only supervision can cause negative transfer. \citet{zhang2025latentshape} reconstruct occupancy fields to pretrain a geometry encoder, then train neural operators using its latent representations. The encoder remains frozen during downstream learning, and the geometry distributions are matched or similar between stages. Their experiments are mainly two-dimensional, alongside a three-dimensional inductor case. These methods motivate using unlabeled geometry collections for pretraining. SimpFlow derives velocity and pressure targets from simplified physical solutions on a cross-category 3D geometry library.

\subsection{Simulation-based pretraining and multi-fidelity transfer}
Simulation-based pretraining learns reusable representations from collections of PDE solutions. MPP~\citep{mccabe2024mpp} shares representations across heterogeneous physical systems, while DPOT~\citep{hao2024dpot} uses autoregressive denoising to stabilize learning from diverse trajectories. Poseidon~\citep{herde2024poseidon} pretrains a multiscale operator on fluid trajectories and adapts it to downstream PDE tasks, including physical systems absent from pretraining. These models rely on simulated trajectories, predominantly represented on regular grids, to support transfer across physical problems.

For CFD, GP-UPT~\citep{bleeker2025neuralcfdv1} demonstrates transfer from DrivAerNet RANS simulations to DrivAerML high-fidelity simulations, while cross-domain pretraining~\citep{zhou2026crossdomain} pools existing 3D CFD datasets across applications. These studies transfer physical information across geometries, operating conditions, and simulation fidelities using existing CFD corpora. SimpFlow constructs its source supervision from an unlabeled geometry library without requiring an existing CFD dataset.

Multi-fidelity learning combines high-fidelity data with cheaper solutions from simplified equations or coarser discretizations~\citep{meng2020multifidelity,lu2022multifidelity}. \citet{setinek2025multifidelity,setinek2026computeoptimal} study data allocation across fidelities, with aerodynamic experiments on two-dimensional airfoils; the latter focuses on discretization and iterative-solver errors. These works examine how simulation fidelity and data allocation affect prediction performance and the cost of obtaining supervision.

Panel-based transfer provides a direct precedent: \citet{wong2025panel} pretrain networks on inviscid solutions for a fixed NACA0012 airfoil and transfer to RANS pressure and skin-friction distributions, including unseen operating conditions. Our setting extends panel-based transfer to pretraining across an unlabeled, cross-category 3D geometry library and adaptation to independently constructed CFD benchmarks.

Concurrent work by \citet{shen2026scaling3d} generates low-fidelity labels on independent 3D source geometries using lattice-Boltzmann simulations and demonstrates downstream transfer. Both approaches use independent source geometries for high-fidelity adaptation. SimpFlow uses a potential-flow solver that requires only surface meshes and reuses precomputed velocity fields across inflow directions to reduce labeling work. We compare label-generation costs in \cref{sec:labelcost}.

\section{Method}
\label{sec:method}

\subsection{Overview}
\begin{figure*}[t]
\centering
\includegraphics[width=\textwidth]{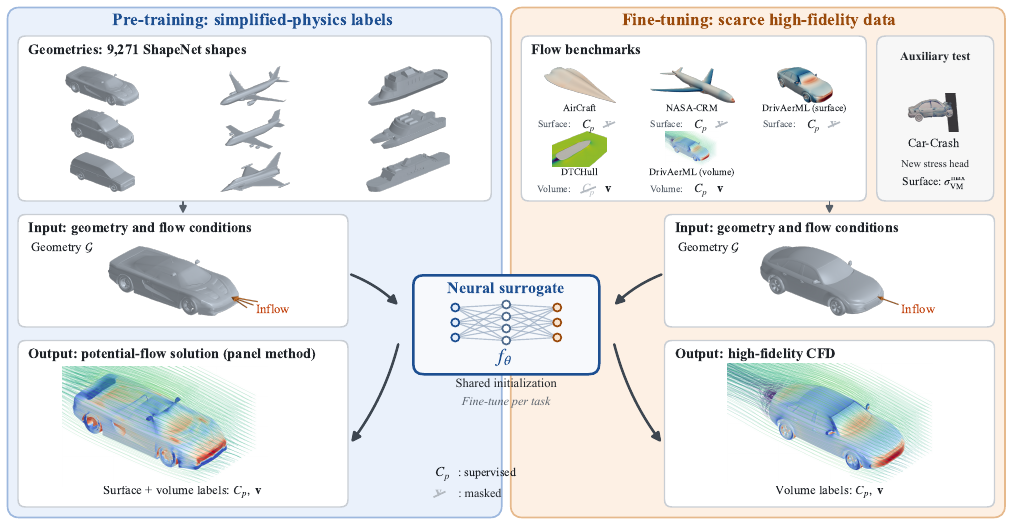}
\caption{SimpFlow pretraining and transfer. Potential-flow labels on 9,271 unlabeled ShapeNet geometries (8,971 training, 300 validation) provide a shared initialization for separate downstream fine-tuning runs. Labels use closed and simplified source meshes; colored flow fields illustrate the references in \cref{fig:physicalfields}. The schematic uses $C_p$ for pressure coefficient, $\mathbf v$ for normalized velocity, and $\sigma_{\mathrm{VM}}^{\max}$ for maximum-over-time von Mises stress; black and struck-through symbols mark supervised and excluded components. Volume DrivAerML supervises all four channels at volume queries, with surface points providing context. The separate Car-Crash test uses a new scalar stress head; its illustration is cropped from GeoPT~\citep{wu2026geopt} (CC BY 4.0).}
\label{fig:method}
\end{figure*}

SimpFlow pretrains a neural surrogate on potential-flow labels from an unlabeled geometry library, then fine-tunes the shared initialization separately on each downstream task and data budget (\cref{fig:method}).

Given geometry $\mathcal G$, conditions $\mathbf c$, and query locations $\mathcal X=\{\mathbf x_i\}_{i=1}^N$, we construct query features $\mathcal Z$ and jointly predict their flow responses:
\begin{equation}
\widehat{\mathbf Y}=f_\theta(\mathcal Z)\in\mathbb R^{N\times4},
\qquad\mathcal Z=\{\mathbf z_i\}_{i=1}^N.
\label{eq:output}
\end{equation}
Both pretraining and downstream flow prediction use dimensionless targets $\mathbf y_i=[C_p(\mathbf x_i),\mathbf v(\mathbf x_i)]$, with
\begin{equation}
C_p=\frac{p-p_\infty}{\tfrac12\rho U_\infty^2},
\qquad\mathbf v=\frac{\mathbf u}{U_\infty},
\label{eq:nondimensional}
\end{equation}
where $p_\infty$, $\rho$, and $U_\infty$ are the freestream pressure, density, and speed. CFD targets use these definitions, with supervision restricted to available components and queries.

For steady, inviscid, incompressible, irrotational flow, Bernoulli's equation gives
\begin{equation}
p+\tfrac12\rho\|\mathbf u\|^2
=p_\infty+\tfrac12\rho U_\infty^2.
\label{eq:bernoulli}
\end{equation}
Substituting into \cref{eq:nondimensional} gives
\begin{equation}
C_p=1-\|\mathbf u\|^2/U_\infty^2=1-\|\mathbf v\|^2.
\label{eq:teacher}
\end{equation}

\subsection{Labeling with Simplified Physics}
Potential flow couples the geometry boundary and permits reuse across inflows. Its dimensionless velocity potential satisfies
\begin{equation}
\begin{aligned}
\Delta\phi&=0 &&\text{in }\Omega_{\mathcal G},\\
\nabla\phi\cdot\mathbf n&=0 &&\text{on }\partial\mathcal G,\\
\nabla\phi&\to\mathbf a &&\text{as }\|\mathbf x\|\to\infty,
\end{aligned}
\label{eq:potential}
\end{equation}
where $\Omega_{\mathcal G}$ is the exterior fluid domain, $\mathbf n$ the wall normal pointing into the fluid, and $\mathbf a=\mathbf U_\infty/U_\infty$ the unit inflow direction. The simplified velocity is $\mathbf v_L=\nabla\phi$, with pressure from \cref{eq:teacher}. The model omits viscosity, separated wakes, turbulence, compressibility, and free-surface effects.

A source-panel method~\citep{hess1967panel,katz2001lowspeed} discretizes only the surface. One matrix factorization yields three velocity bases $\mathbf b_{\mathcal G,k}$ for Cartesian unit inflows $\mathbf e_k$, giving
% Keep the figure-heavy method page from stretching display spacing.
\raggedbottom
\begingroup
\setlength{\abovedisplayskip}{7pt}
\setlength{\belowdisplayskip}{7pt}
\setlength{\abovedisplayshortskip}{4pt}
\setlength{\belowdisplayshortskip}{7pt}
\begin{equation}
\mathbf v_L(\mathbf x;\mathbf a)
=\sum_{k=1}^{3}a_k\,\mathbf b_{\mathcal G,k}(\mathbf x).
\label{eq:inflowbasis}
\end{equation}
\endgroup
New query clouds require basis evaluation but no new source solve. At fixed queries, new inflows reuse these values; pressure is recomputed through \cref{eq:teacher}. Solver details and costs are in \cref{app:teacher,app:cost}.

Source geometries are ShapeNet cars, airplanes, and watercraft~\citep{chang2015shapenet}, aligned in orientation and scaled to streamwise length five. We construct closed surfaces for sampling and geometric queries and simplify them to a target of 4,000 panels for the main solver. Each sample pairs a geometry and inflow with 4,096 surface and 8,192 volume queries, with volume points sampled both near the wall and farther out. Each geometry provides 40 samples: cars vary query clouds at fixed inflow, while airplanes and watercraft also vary inflow conditions (\cref{app:sampling}).

Surface targets use the nearest panel's fluid-side center velocity, while volume fields are evaluated at query locations. Validity masking and label clipping follow \cref{app:teacher}.

\subsection{Training and Transfer}
The features in \cref{eq:output} encode query position, geometry, and conditions:
\begingroup
\setlength{\abovedisplayskip}{7pt}
\setlength{\belowdisplayskip}{7pt}
\setlength{\abovedisplayshortskip}{4pt}
\setlength{\belowdisplayshortskip}{7pt}
\begin{equation}
\mathbf z_i=[\mathbf x_i,d_i,\boldsymbol\eta_i,\mathbf a,a_{c,i}]
\in\mathbb R^{11}.
\label{eq:input}
\end{equation}
\endgroup
For volume queries, $d_i$ is the distance to the nearest surface point and $\boldsymbol\eta_i$ is the unit vector pointing toward it. For surface queries, $d_i=0$ and $\boldsymbol\eta_i$ is the outward unit normal. The conditions $\mathbf c$ are encoded by the shared inflow direction $\mathbf a$ and a scalar $a_{c,i}$ representing a Mach code or fluid-region indicator. These encodings follow GeoPT~\citep{wu2026geopt} and are specified in \cref{app:interface}. The potential-flow teacher depends on geometry and inflow direction; the Mach code does not introduce compressible physics into its labels.

We use Transolver~\citep{wu2024transolver} to process $\mathcal Z$ jointly. It aggregates queries into learned slice tokens, applies attention among the tokens, and broadcasts the result back to the queries. Its output head returns the four channels in \cref{eq:output}.

Both stages minimize masked mean squared error. For a batch $\mathcal B$, let $\mathcal O_{\mathcal B}$ contain its valid, observed sample--query--component indices. With $\widehat{\mathbf Y}_b=f_\theta(\mathcal Z_b)$, the loss is
\begin{equation}
\mathcal L_{\mathcal B}(\theta)=\frac{1}{|\mathcal O_{\mathcal B}|}
\sum_{(b,i,k)\in\mathcal O_{\mathcal B}}
\big(\widehat Y_{b,i,k}-Y_{b,i,k}\big)^2.
\label{eq:loss}
\end{equation}
\flushbottom
Pretraining supervises all four components at valid queries using the postprocessed potential-flow targets. Fine-tuning includes only the quantities and query points observed by each task; points used solely as geometry context receive no loss (\cref{app:experiments}).

Let $\theta_L$ denote the pretrained weights. Each flow task starts from $\theta_L$ and updates all parameters, including the four-channel output head. The separate Car-Crash test inherits the embedding and backbone and initializes a new scalar stress head.

\subsection{Local Geometry Field Control}
\label{sec:lgf}
Local Geometry Field (LGF) constructs velocity from local geometry, using the same inputs, targets, and training schedule as SimpFlow (validity masks: \cref{app:teacher}). For unit $\mathbf a$ and $\boldsymbol\eta$,
\begin{equation}
\begin{aligned}
\mathbf v_s&=\mathbf a-\omega(d)(\mathbf a\cdot\boldsymbol\eta)\boldsymbol\eta,\\
\omega(d)&=\exp[-(d/\delta)^2].
\end{aligned}
\label{eq:smooth}
\end{equation}
with $\delta=0.5$ in normalized geometry coordinates.
\begin{figure}[!ht]
\centering
\includegraphics[width=\columnwidth]{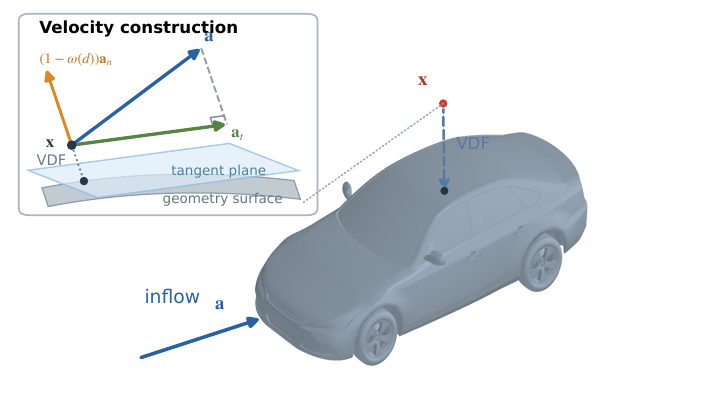}
\caption{LGF construction at an exterior query $\mathbf x$. The inset shows the tangent plane at the nearest surface point, where the inflow splits into a kept tangential part and a damped normal part: $\mathbf v_s=\mathbf a_t+(1-\omega(d))\mathbf a_n$ with $\mathbf a_n=(\mathbf a\cdot\boldsymbol\eta)\boldsymbol\eta$.}
\label{fig:smooth}
\end{figure}
LGF preserves tangential inflow, damps normal inflow near the wall, and recovers the freestream far away (\cref{fig:smooth}). Its pressure coefficient follows from \cref{eq:teacher}:
\begin{equation}
C_{p,s}=1-\|\mathbf v_s\|^2
=\omega(2-\omega)(\mathbf a\cdot\boldsymbol\eta)^2\ge0.
\label{eq:smoothcp}
\end{equation}
LGF therefore cannot produce suction; the derivation is in \cref{app:smooth}.

\FloatBarrier

\section{Experiments}
\label{sec:experiments}
We evaluate labeling cost and transfer to five flow tasks, analyze the effects of supervision content and panel resolution, and finally examine transfer to solid mechanics.

\paragraph{Setup.} We evaluate surface $C_p$ on AirCraft~\citep{luo2025transolverpp}, NASA-CRM~\citep{bekemeyer2025crm}, and DrivAerML~\citep{ashton2024drivaerml}, and normalized volume velocity $\mathbf v$ on DTCHull~\citep{wu2026geopt}. DrivAerML volume predicts all four components $[C_p,\mathbf v]$ at volume queries; surface context points receive no loss. Flow targets use the nondimensionalization in \cref{eq:nondimensional}. The original four tasks use 20/40/60/80/100 training cases; volume DrivAerML uses 10/20/30/40, defining 24 flow task--budget settings. Counts refer to complete CFD cases. Car-Crash provides a separate solid-mechanics test.

We compare Scratch, LGF, GeoPT~\citep{wu2026geopt}, and SimpFlow, using an 8-layer Transolver and 200 fine-tuning epochs. Pretrained methods use 8,971 source geometries and 358,840 training examples; SimpFlow's main initialization uses 4k-panel labels. All main runs use seed 42 for pretraining and fine-tuning, fixed task-specific splits, and shared backbone and optimization settings. GeoPT retains its input mapping and resets its output head; SimpFlow and LGF retain both for flow prediction. GeoPT is retrained at the shared data scale (\cref{app:reproductionscope}); its separate early reproduction and input-mapping control are in \cref{app:earlygeopt}. We report test relative $L_2$ averaged over cases, with final comparisons at epoch 200. These are single-seed results. Full protocols are in \cref{app:experiments}.

\begin{figure*}[t]
\centering
\includegraphics[width=\textwidth]{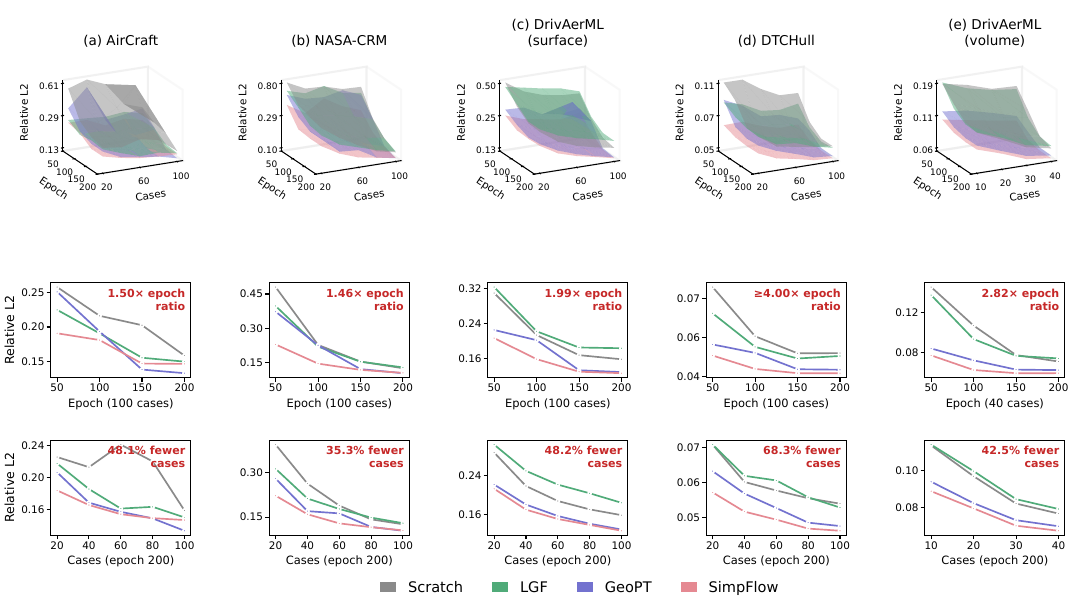}
\caption{Test relative $L_2$ across five flow benchmarks, using seed 42 and 4k-panel pretraining for SimpFlow. Top: observed budget--epoch grids; facets connect measured values without fitting. Vertical axes are logarithmic and labeled with actual errors. Middle: epoch curves at each task's largest budget (100/100/100/100/40 cases). Bottom: budget curves at epoch 200, including 10/20/30/40 cases for volume DrivAerML. Two-dimensional axes are linear. Values are case means from one seed, without uncertainty shading. Red annotations give interpolated case savings at epoch 200 and epoch ratios at the largest budget, relative to scratch's final-budget, epoch-200 error (\cref{app:efficiency}). Epoch ratios exclude pretraining and labeling costs and do not measure wall-clock speedup. Targets are surface $C_p$ for AirCraft, NASA-CRM, and DrivAerML, volume $\mathbf v$ for DTCHull, and volume $[C_p,\mathbf v]$ for DrivAerML. Car-Crash is analyzed separately in \cref{sec:crashtransfer}. Full grids and curves are in \cref{app:additionalresults}.}
\label{fig:transfer}
\end{figure*}

\subsection{Labeling Cost and Flow Transfer}
\label{sec:main}
\label{sec:labelcost}
\paragraph{Low-cost physical labels.} With four logical CPUs of an AMD EPYC 9684X, SimpFlow takes 0.111\,s per example, 64.2\% less than our GeoPT reproduction (\cref{tab:threewaycost}). The benchmark uses one geometry per category, four query clouds and ten conditions per cloud, totaling 120 examples. Timing covers labeling from existing meshes and queries, including validity checks, but excludes mesh preparation, query sampling, and I/O. Factorization and velocity-basis reuse amortize work across conditions; the actual car corpus varies clouds at fixed inflow (\cref{app:cost}).

\begin{table}[!ht]
\centering\maintablestyle
\caption{Amortized labeling time per benchmark example. SimpFlow and our GeoPT reproduction use the same CPU allocation over 120 examples. LBM is estimated from the reported 542\,s geometry--condition solve assuming four query clouds per solution; extraction is omitted. Accounting is in \cref{app:cost}.}
\label{tab:threewaycost}
\begin{tabular}{llr}
\toprule
\textbf{Method} & \textbf{Hardware} & \textbf{s/example}\\
\midrule
SimpFlow (4k) & EPYC 9684X, 4 threads & \textbf{0.111}\\
\shortstack[l]{GeoPT\\\citep{wu2026geopt}} & EPYC 9684X, 4 threads & 0.309\\
\shortstack[l]{LBM\\\citep{shen2026scaling3d}} & NVIDIA A100, 1 GPU & 135.5$^{*}$\\
\bottomrule
\end{tabular}
\par\vspace{2pt}\raggedright\footnotesize
$^{*}$Reported solve time divided by four, rather than a new measurement: $542/4=135.5$\,s.
\end{table}

The concurrent LBM study reports 542\,s per geometry--condition solve on an A100~\citep{shen2026scaling3d}. Sharing each solution across four clouds gives a solve-only estimate of 135.5\,s per example, about 1,225 times SimpFlow's measured mean. This hardware-specific comparison concerns labeling cost; transfer comparisons below use baselines reproduced under a shared training protocol.

\begin{figure*}[!t]
\centering
\includegraphics[width=\textwidth]{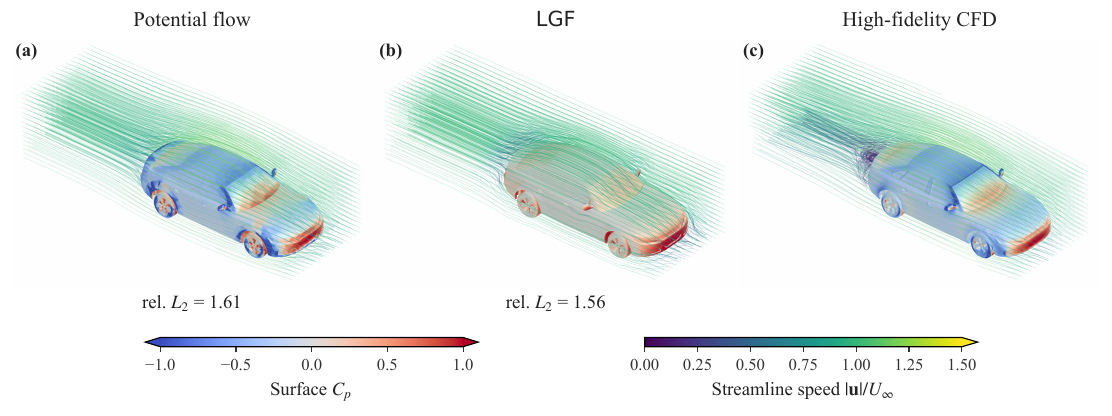}
\caption{Reference fields for DrivAerML run\_181 with 8,000-panel potential flow. The surface $C_p$ color scale is clipped to $[-1,1]$ and normalized streamline speed to $[0,1.5]$. Potential flow and CFD share the stagnation high-pressure and roof low-pressure regions, whereas only CFD has a separated wake. Clipping hides amplitude differences, so the figure compares structure. Large annotated errors use identical full surface queries before display averaging or clipping. Dataset and per-sample errors, including exclusion-based diagnostics, are in \cref{app:referenceaccuracy}; further views are in \cref{fig:physicaldetails,fig:aircraftreference}.}
\label{fig:physicalfields}
\end{figure*}

\paragraph{Consistent flow transfer.} At epoch 200, the shared SimpFlow initialization improves over scratch in all 24 settings and GeoPT in 21 (\cref{fig:transfer}). It beats GeoPT at every surface DrivAerML, DTCHull, and volume DrivAerML budget, extending the benefit from individual flow quantities to four-channel prediction. GeoPT performs better on AirCraft at 80 and 100 cases and NASA-CRM at 100 cases. Full errors are in \cref{app:learninggrid}; showcases use model-selection checkpoints (\cref{app:showcases}).

\paragraph{Fewer CFD cases and fine-tuning epochs.} We use scratch's largest-budget, epoch-200 error as the task-specific reference. At epoch 200, intersections with the budget curves give 35--68\% fewer required CFD cases; at the largest budgets, epoch ratios range from 1.46 to at least 4.00 (\cref{fig:transfer}). DTCHull reaches the reference by the first measured checkpoint, epoch 50. These interpolated comparisons measure fine-tuning data and epochs, excluding pretraining and labeling costs. Per-task values and calculations are in \cref{app:efficiency}, with training histories in \cref{app:learningcurves}.

\subsection{Effect of Pretraining Supervision}
\label{sec:supervision}
\label{sec:why}

LGF and SimpFlow share the geometry library, prediction targets, and training schedule, and both retain the output head (validity masks: \cref{app:teacher}). Yet LGF causes negative transfer at every surface DrivAerML budget, while SimpFlow improves over scratch in all flow settings and reduces error relative to LGF by 20.7\% on average.

Potential flow couples the entire boundary, allowing geometry-induced acceleration and deceleration. Through \cref{eq:teacher}, these produce suction and high-pressure regions. LGF only attenuates inflow, so $\|\mathbf v_s\|\leq1$ and $C_{p,s}\geq0$ (derivation and sphere comparison: \cref{app:smooth}). Despite similar-looking streamlines, it misses the roof suction shared by potential flow and CFD (\cref{fig:physicalfields}).

Potential flow retains frontal stagnation pressure and roof suction but misses the separated wake. Its surface relative $L_2$ on this car is 1.61, versus 1.56 for LGF; these reference fields use 8k panels, while transfer uses 4k. Thus, useful flow structure can accompany large pointwise errors. Further fields and label errors are in \cref{fig:aircraftreference,fig:hullreference,app:referenceaccuracy}.

Resetting the main 4k output head still beats scratch in 24/24 settings and GeoPT in 21/24 (\cref{app:headreset}), showing that transfer gains do not depend on head inheritance.

\FloatBarrier
\subsection{Effect of Panel Resolution}
\label{sec:panelresolution}
With the physical model fixed, we next ask whether finer boundary resolution continually improves transfer. We vary the target panel count from 500 to 8k, holding geometry identities, inflows, query clouds, initialization, and optimization schedules fixed. Each model is fine-tuned on 20 CFD cases per benchmark and evaluated at epoch 200. The dedicated 8k run shares the 2k/4k validity mask (\cref{app:ablations}).

\begin{figure}[!ht]
\centering
\includegraphics[width=\columnwidth]{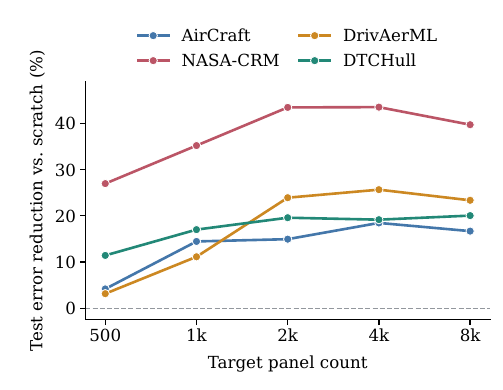}
\caption{Downstream test-error reduction relative to scratch versus target panel count. Each curve represents one benchmark; points use 20 CFD training cases, seed 42, and epoch 200. Higher is better; the dashed line denotes no gain. Numerical results are listed in \cref{tab:panelresolution}.}
\label{fig:panelresolution}
\end{figure}

Very coarse resolution weakens transfer: reducing the panel count from 1k to 500 lowers the gain over scratch by 5.6--10.3 percentage points across the four tasks, although both improve over scratch (\cref{fig:panelresolution}). From 2k to 8k, finer resolution gives no consistent monotonic improvement. The lowest errors occur at 4k on three tasks and 8k on DTCHull (\cref{tab:panelresolution}), indicating that moderate resolution already supplies useful supervision.

The broader comparison uses 22 shared task--budget settings, including volume DrivAerML at 20/40 cases; its 10/30-case runs are available only for 4k. Wins over GeoPT increase from 15/22 at 2k to 19/22 at 4k and remain 19/22 at 8k (\cref{tab:resolutionwins}). This comparison uses the original 8k initialization and mask, distinct from the dedicated resolution study. Meanwhile, labeling costs rise from 0.055 to 0.111 and 0.322\,s per example. These results motivate 4k as the main configuration: it balances transfer performance and labeling cost.

\begin{table}[!ht]
\centering\maintablestyle
\caption{Settings with lower epoch-200 test error than GeoPT. Counts use the same 22 settings across five flow tasks: five budgets for each original task and 20/40 cases for volume DrivAerML. The 8k column uses the original initialization and validity mask; the 20-case curves use dedicated resolution-study runs.}
\label{tab:resolutionwins}
\begin{tabular}{lrrr}
\toprule
\textbf{Task} & \textbf{2k} & \textbf{4k} & \textbf{8k}\\
\midrule
AirCraft & 3/5 & 3/5 & 4/5\\
NASA-CRM & 3/5 & 4/5 & 3/5\\
DrivAerML (surface) & 2/5 & 5/5 & 5/5\\
DTCHull & 5/5 & 5/5 & 5/5\\
DrivAerML (volume) & 2/2 & 2/2 & 2/2\\
\midrule
Total & 15/22 & 19/22 & 19/22\\
\bottomrule
\end{tabular}
\end{table}

\FloatBarrier
\subsection{Transfer to Solid Mechanics}
\label{sec:crashtransfer}
We finally test whether a flow-pretrained representation can help predict structural response. Car-Crash takes an undeformed car geometry as input and predicts maximum-over-time von Mises stress. All pretrained methods replace the output layer with a scalar stress head and use 20, 40, 60, or 80 training cases, following the same 200-epoch schedule and seed 42.

\begin{figure}[!ht]
\centering
\includegraphics[width=\linewidth]{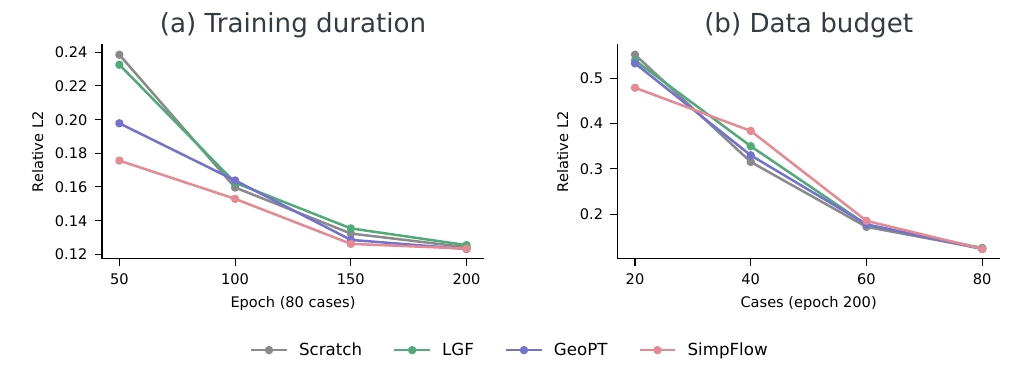}
\includegraphics[width=.88\linewidth]{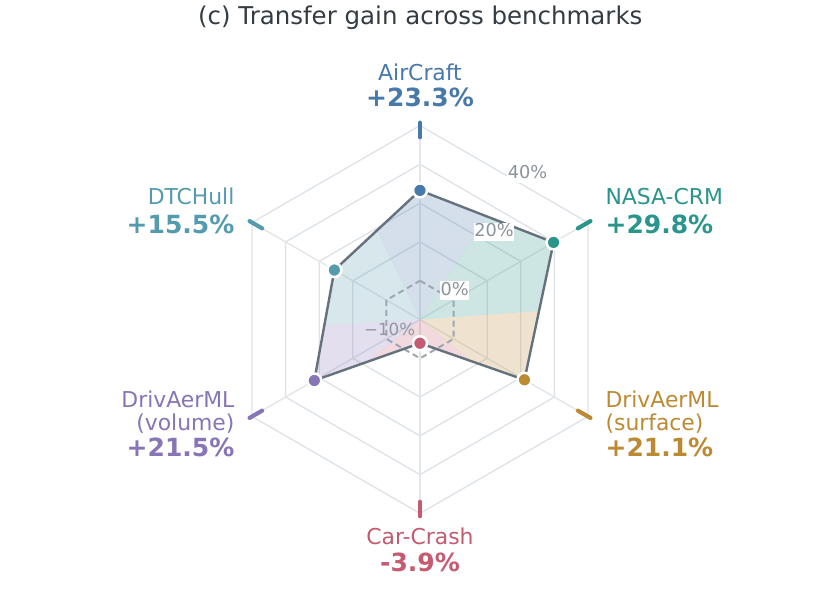}
\caption{Transfer with the same 4k initialization. Car-Crash test relative $L_2$ versus (a) epoch at 80 cases and (b) budget at epoch 200. (c) Mean relative error reduction against scratch across all evaluated budgets of each benchmark at epoch 200, computed as $100(1-E_{\mathrm{SimpFlow}}/E_{\mathrm{Scratch}})$ per budget and then averaged equally. All six axes use the same $[-10,40]\%$ scale; the dashed hexagon marks zero gain. Full results are in \cref{app:learninggrid,app:crashresults}.}
\label{fig:crashtransfer}
\end{figure}

At 20 training cases, SimpFlow reduces test error by 13.2\% relative to scratch and also outperforms GeoPT (\cref{fig:crashtransfer}). At 40 and 60 cases its error is 21.7\% and 7.7\% above scratch, respectively, while at 80 cases it is 0.7\% below scratch. Averaged across the evaluated budgets, error reduction is 15.5--29.8\% on the five flow tasks but $-3.9\%$ on Car-Crash (\cref{fig:crashtransfer}c). Transfer is thus stronger and more consistent for flow prediction than for structural response. This difference is consistent with the physical content of pretraining supervision: potential flow links geometry to flow response, whereas crash-induced stress also depends on distinct mechanisms such as material behavior, contact, and structural deformation. The pressure and stress examples in \cref{fig:crashreference} illustrate the change in physical response; complete results are in \cref{app:crashresults}.

\FloatBarrier
\section{Conclusion}
SimpFlow turns unlabeled 3D geometry libraries into physical supervision for transferable flow representations, without existing CFD data or paired downstream solutions. Boundary-only potential-flow solves and reusable inflow bases enable inexpensive labeling. The resulting initialization transfers to independent CFD benchmarks, improving over scratch in all 24 tested settings and reducing interpolated CFD sample requirements by 35--68\% at a fixed fine-tuning budget.

The results show that simplified solutions can provide useful supervision despite large pointwise errors. Potential flow retains geometry-induced acceleration, stagnation, and suction that the local attenuation rule of LGF cannot fully express. Moderate panel resolution provides a favorable cost--transfer balance, while further refinement yields no consistent monotonic improvement. SimpFlow thus offers a practical route to labeling extensive geometry libraries for pretraining and reducing the high-fidelity data demands of CFD surrogates.

\bibliography{references}

@article{li2024pino,
  title   = {Physics-Informed Neural Operator for Learning Partial Differential Equations},
  author  = {Li, Zongyi and Zheng, Hongkai and Kovachki, Nikola and Jin, David and Chen, Haoxuan and Liu, Burigede and Azizzadenesheli, Kamyar and Anandkumar, Anima},
  journal = {ACM/IMS Journal of Data Science},
  year    = {2024},
  doi     = {10.1145/3648506}
}

@article{raissi2019pinn,
  title   = {Physics-informed neural networks: A deep learning framework for solving forward and inverse problems involving nonlinear partial differential equations},
  author  = {Raissi, Maziar and Perdikaris, Paris and Karniadakis, George E.},
  journal = {Journal of Computational Physics},
  volume  = {378},
  pages   = {686--707},
  year    = {2019}
}

@article{karniadakis2021piml,
  title   = {Physics-informed machine learning},
  author  = {Karniadakis, George Em and Kevrekidis, Ioannis G. and Lu, Lu and Perdikaris, Paris and Wang, Sifan and Yang, Liu},
  journal = {Nature Reviews Physics},
  volume  = {3},
  number  = {6},
  pages   = {422--440},
  year    = {2021}
}

@article{lu2021deeponet,
  title   = {Learning nonlinear operators via {DeepONet} based on the universal approximation theorem of operators},
  author  = {Lu, Lu and Jin, Pengzhan and Pang, Guofei and Zhang, Zhongqiang and Karniadakis, George Em},
  journal = {Nature Machine Intelligence},
  volume  = {3},
  number  = {3},
  pages   = {218--229},
  year    = {2021}
}

@inproceedings{li2021fno,
  title     = {Fourier neural operator for parametric partial differential equations},
  author    = {Li, Zongyi and Kovachki, Nikola Borislavov and Azizzadenesheli, Kamyar and Liu, Burigede and Bhattacharya, Kaushik and Stuart, Andrew and Anandkumar, Anima},
  booktitle = {International Conference on Learning Representations},
  year      = {2021}
}

@inproceedings{li2023gino,
  title     = {Geometry-informed neural operator for large-scale {3D} {PDEs}},
  author    = {Li, Zongyi and Kovachki, Nikola Borislavov and Choy, Chris and Li, Boyi and Kossaifi, Jean and Otta, Shourya Prakash and Nabian, Mohammad Amin and Stadler, Maximilian and Hundt, Christian and Azizzadenesheli, Kamyar and Anandkumar, Anima},
  booktitle = {Advances in Neural Information Processing Systems},
  year      = {2023}
}

@inproceedings{hao2023gnot,
  title     = {{GNOT}: A general neural operator transformer for operator learning},
  author    = {Hao, Zhongkai and Wang, Zhengyi and Su, Hang and Ying, Chengyang and Dong, Yinpeng and Liu, Songming and Cheng, Ze and Song, Jian and Zhu, Jun},
  booktitle = {International Conference on Machine Learning},
  year      = {2023}
}

@inproceedings{wu2024transolver,
  title     = {Transolver: A fast transformer solver for {PDEs} on general geometries},
  author    = {Wu, Haixu and Luo, Huakun and Wang, Haowen and Wang, Jianmin and Long, Mingsheng},
  booktitle = {International Conference on Machine Learning},
  year      = {2024}
}

@inproceedings{luo2025transolverpp,
  title     = {Transolver++: An accurate neural solver for {PDEs} on million-scale geometries},
  author    = {Luo, Huakun and Wu, Haixu and Zhou, Hang and Xing, Lanxiang and Di, Yichen and Wang, Jianmin and Long, Mingsheng},
  booktitle = {International Conference on Machine Learning},
  year      = {2025}
}

@inproceedings{alkin2024upt,
  title     = {Universal physics transformers: A framework for efficiently scaling neural operators},
  author    = {Alkin, Benedikt and F{\"u}rst, Andreas and Schmid, Simon and Gruber, Lukas and Holzleitner, Markus and Brandstetter, Johannes},
  booktitle = {Advances in Neural Information Processing Systems},
  year      = {2024}
}

@article{alkin2025abupt,
  title   = {{AB-UPT}: Scaling neural {CFD} surrogates for high-fidelity automotive aerodynamics simulations via anchored-branched universal physics transformers},
  author  = {Alkin, Benedikt and Bleeker, Maurits and Kurle, Richard and Kronlachner, Tobias and Sonnleitner, Reinhard and Dorfer, Matthias and Brandstetter, Johannes},
  journal = {Transactions on Machine Learning Research},
  year    = {2025}
}

@article{ashton2024drivaerml,
  title   = {{DrivAerML}: High-fidelity computational fluid dynamics dataset for road-car external aerodynamics},
  author  = {Ashton, Neil and Mockett, Charles and Fuchs, Marian and Fliessbach, Louis and Hetmann, Hendrik and Knacke, Thilo and Schonwald, Norbert and Skaperdas, Vangelis and Fotiadis, Grigoris and Walle, Astrid and others},
  journal = {arXiv preprint arXiv:2408.11969},
  year    = {2024}
}

@inproceedings{bekemeyer2025crm,
  title     = {Introduction of applied aerodynamics surrogate modeling benchmark cases},
  author    = {Bekemeyer, Philipp and Hariharan, Nathan and Wissink, Andrew M. and Cornelius, Jason},
  booktitle = {AIAA SCITECH 2025 Forum},
  year      = {2025}
}

@inproceedings{mccabe2024mpp,
  title     = {Multiple physics pretraining for spatiotemporal surrogate models},
  author    = {McCabe, Michael and R{\'e}galdo-Saint Blancard, Bruno and Parker, Liam Holden and Ohana, Ruben and Cranmer, Miles and Bietti, Alberto and Eickenberg, Michael and Golkar, Siavash and Krawezik, Geraud and Lanusse, Francois and Pettee, Mariel and Tesileanu, Tiberiu and Cho, Kyunghyun and Ho, Shirley},
  booktitle = {Advances in Neural Information Processing Systems},
  year      = {2024}
}

@inproceedings{hao2024dpot,
  title     = {{DPOT}: Auto-regressive denoising operator transformer for large-scale {PDE} pre-training},
  author    = {Hao, Zhongkai and Su, Chang and Liu, Songming and Berner, Julius and Ying, Chengyang and Su, Hang and Anandkumar, Anima and Song, Jian and Zhu, Jun},
  booktitle = {International Conference on Machine Learning},
  year      = {2024}
}

@inproceedings{herde2024poseidon,
  title     = {Poseidon: Efficient foundation models for {PDEs}},
  author    = {Herde, Maximilian and Raoni{\'c}, Bogdan and Rohner, Tobias and K{\"a}ppeli, Roger and Molinaro, Roberto and de B{\'e}zenac, Emmanuel and Mishra, Siddhartha},
  booktitle = {Advances in Neural Information Processing Systems},
  year      = {2024}
}

@article{meng2020multifidelity,
  title   = {A composite neural network that learns from multi-fidelity data: Application to function approximation and inverse {PDE} problems},
  author  = {Meng, Xuhui and Karniadakis, George Em},
  journal = {Journal of Computational Physics},
  volume  = {401},
  pages   = {109020},
  year    = {2020}
}

@article{lu2022multifidelity,
  title   = {Multifidelity deep neural operators for efficient learning of partial differential equations with application to fast inverse design of nanoscale heat transport},
  author  = {Lu, Lu and Pestourie, Rapha{\"e}l and Johnson, Steven G. and Romano, Giuseppe},
  journal = {Physical Review Research},
  volume  = {4},
  number  = {2},
  pages   = {023210},
  year    = {2022}
}

@article{wong2025panel,
  title   = {Inductive transfer-learning of high-fidelity aerodynamics from inviscid panel methods},
  author  = {Wong, Benjamin Y. J. and Khoo, Boo Cheong},
  journal = {Advances in Aerodynamics},
  year    = {2025},
  volume  = {7},
  pages   = {1},
  doi     = {10.1186/s42774-024-00186-0}
}

@inproceedings{wu2026geopt,
  title     = {{GeoPT}: Scaling physics simulation via lifted geometric pre-training},
  author    = {Wu, Haixu and Guo, Minghao and Li, Zongyi and Dou, Zhiyang and Long, Mingsheng and He, Kaiming and Matusik, Wojciech},
  booktitle = {International Conference on Machine Learning},
  year      = {2026},
  eprint = {2602.20399},
  archivePrefix = {arXiv}
}

@article{chang2015shapenet,
  title   = {{ShapeNet}: An information-rich {3D} model repository},
  author  = {Chang, Angel X. and Funkhouser, Thomas and Guibas, Leonidas and Hanrahan, Pat and Huang, Qixing and Li, Zimo and Savarese, Silvio and Savva, Manolis and Song, Shuran and Su, Hao and others},
  journal = {arXiv preprint arXiv:1512.03012},
  year    = {2015}
}

@article{hess1967panel,
  title   = {Calculation of potential flow about arbitrary bodies},
  author  = {Hess, John L. and Smith, A. M. O.},
  journal = {Progress in Aerospace Sciences},
  volume  = {8},
  pages   = {1--138},
  year    = {1967}
}

@book{katz2001lowspeed,
  title     = {Low-Speed Aerodynamics},
  author    = {Katz, Joseph and Plotkin, Allen},
  edition   = {2},
  publisher = {Cambridge University Press},
  year      = {2001}
}

@misc{zhou2026crossdomain,
      title={Cross-Domain Pretraining for Steady-State Neural {CFD} Surrogates}, 
      author={Anthony Zhou and Amir Barati Farimani and Shirley Ho and Rudy Morel},
      year={2026},
      eprint={2610.10398},
      archivePrefix={arXiv},
      primaryClass={cs.LG},
      url={https://arxiv.org/abs/2610.10398}, 
      howpublished={arXiv preprint arXiv:2610.10398},
}

@misc{bleeker2025neuralcfdv1,
      title={{NeuralCFD}: Deep Learning on High-Fidelity Automotive Aerodynamics Simulations}, 
      author={Maurits Bleeker and Matthias Dorfer and Tobias Kronlachner and Reinhard Sonnleitner and Benedikt Alkin and Johannes Brandstetter},
      year={2025},
      eprint={2502.09692},
      archivePrefix={arXiv},
      primaryClass={cs.LG},
      url={https://arxiv.org/abs/2502.09692v1}, 
      note={Version 1},
      howpublished={arXiv preprint arXiv:2502.09692v1},
}

@misc{setinek2025multifidelity,
      title={Towards Multi-Fidelity Scaling Laws of Neural Surrogates in {CFD}}, 
      author={Paul Setinek and Gianluca Galletti and Johannes Brandstetter},
      year={2025},
      eprint={2511.01830},
      archivePrefix={arXiv},
      primaryClass={cs.LG},
      url={https://arxiv.org/abs/2511.01830}, 
      howpublished={arXiv preprint arXiv:2511.01830},
}

@misc{setinek2026computeoptimal,
      title={Compute-optimal data scaling for neural surrogates via multi-fidelity training}, 
      author={Paul Setinek and Fabian Paischer and Nils Thuerey and Johannes Brandstetter and Felix Koehler},
      year={2026},
      eprint={2610.17221},
      archivePrefix={arXiv},
      primaryClass={cs.LG},
      url={https://arxiv.org/abs/2610.17221}, 
      howpublished={arXiv preprint arXiv:2610.17221},
}

@misc{zhang2025latentshape,
      title={From Cheap Geometry to Expensive Physics: Elevating Neural Operators via Latent Shape Pretraining}, 
      author={Zhizhou Zhang and Youjia Wu and Kaixuan Zhang and Yanjia Wang},
      year={2025},
      eprint={2509.25788},
      archivePrefix={arXiv},
      primaryClass={cs.LG},
      url={https://arxiv.org/abs/2509.25788}, 
      howpublished={arXiv preprint arXiv:2509.25788},
}

@misc{shen2026scaling3d,
  title = {Scaling {3D} Physics Surrogates: Low-Fidelity Pretraining for High-Fidelity Transfer},
  author = {Shen, Qianfan and Liu, Yu and Wang, Hao},
  year = {2026},
  howpublished = {Manuscript}
}
\bibliographystyle{icml2026}

\clearpage
\appendix
\onecolumn
\raggedbottom
% Top-align appendix float pages instead of stretching the gaps between tables.
\makeatletter
\setlength{\@fptop}{0pt}
\setlength{\@fpsep}{12pt plus 2pt minus 2pt}
\setlength{\@fpbot}{0pt plus 1fil}
\makeatother
\section{Pre-Training Sample Generation}
\label{app:teacher}

\subsection{Potential-Flow Solver}
Let a closed surface have $P$ triangular panels with constant source strengths $\sigma_j$. Using an outward source-velocity kernel, we write
\begin{equation}
\mathbf K_j(\mathbf x)=\frac{1}{4\pi}\int_{\Gamma_j}
\frac{\mathbf x-\mathbf y}{\|\mathbf x-\mathbf y\|^3}\,\mathrm dS_{\mathbf y},
\quad \mathbf v(\mathbf x)=\mathbf a+\sum_j\sigma_j\mathbf K_j(\mathbf x).
\end{equation}
Collocating the exterior no-penetration condition at panel centers gives
\begin{equation}
A\boldsymbol\sigma=-N\mathbf a,\qquad
A_{ij}=\mathbf n_i\cdot\mathbf K_j^+(\mathbf c_i),
\end{equation}
where row $i$ of $N$ is $\mathbf n_i^\top$. The exterior source-layer jump contributes $+\tfrac12\sigma_i\mathbf n_i$, giving diagonal entries $1/2$ in the normal system. Surface velocity evaluation also retains the in-plane self-interaction of irregular triangles. This source-only model includes neither wake circulation nor a Kutta condition.
Three inflow bases share the same factorization:
\begin{equation}
A\Sigma=-N,\quad B(\mathbf x)=I_3+K(\mathbf x)\Sigma,
\quad\mathbf v(\mathbf x;\mathbf a)=B(\mathbf x)\mathbf a.
\end{equation}
Pressure is computed after combining velocities; pressure fields do not combine linearly. New query clouds require new field evaluations, while the geometry-dependent matrix factorization remains reusable.

For panel $j$ of area $S_j$, we use centroid quadrature when the query-to-centroid distance is at least $3\sqrt{S_j}$. Closer interactions use the centroids of 64 equal-area subtriangles obtained by three subdivision steps; self-interactions use the terms above. Source strengths are solved in double precision. Volume velocities are evaluated at the query coordinates. For each surface query, a closest-surface search identifies the nearest triangle, and its fluid-side panel-center velocity is assigned to that query. All four target components at volume queries inside the panel mesh are marked invalid and excluded from training; the retained LGF control uses the original 8k mask, while the main 4k model uses the shared resolution-study mask (\cref{app:trainingconfig}). To limit extreme label values, we clip pressure coefficients and cap velocity magnitudes:
\begin{equation}
\widetilde{\mathbf v}=\mathbf v\min(1,\sqrt3/\|\mathbf v\|),\qquad
\widetilde C_p=\operatorname{clip}(1-\|\mathbf v\|^2,-2,2).
\end{equation}
For these potential-flow targets, the paired operation preserves $\widetilde C_p=1-\|\widetilde{\mathbf v}\|^2$ up to rounding. Spatially varying rescaling need not preserve the continuous divergence or curl constraints.

\subsection{Sample Construction}
\label{app:sampling}
ShapeNet geometries are centered, oriented by category to match the downstream coordinate conventions, and scaled to longitudinal length five. Preprocessing produces a closed surface for sampling and geometric queries, and a simplified mesh for the source-panel solve. The main experiments target 4,000 panels using quadric edge-collapse decimation with topology and normal preservation. The actual retained count is recorded; meshes with fewer faces or incomplete decimation are used without forcing the target count.

A sample consists of one geometry, one inflow condition, and one query cloud. \Cref{tab:pretrainpoints,tab:categorysampling} specify the point sets and category-dependent sampling. Interior volume candidates are rejected. Potential-flow labels that become invalid under the simplified panel geometry are masked as described above.

\begin{table}[!ht]
\centering\appendixtablestyle
\caption{Query points per pre-training sample. Distances use normalized geometry coordinates.}
\label{tab:pretrainpoints}
\begin{tabular}{lrp{0.65\linewidth}}
\toprule
\textbf{Point set} & \textbf{Count} & \textbf{Sampling rule}\\
\midrule
Surface & 4,096 & Sample triangle locations in proportion to surface area.\\
Near-wall volume & 4,096 & Offset surface samples along outward normals, with log-uniform distance in $[0.02,0.5]$; retain exterior points.\\
Exterior volume & 4,096 & Sample uniformly within the bounding box padded by one unit; retain exterior points.\\
\midrule
Total & 12,288 & One fixed cloud shared by the conditions assigned to it.\\
\bottomrule
\end{tabular}
\end{table}

\begin{table}[!ht]
\centering\appendixtablestyle
\caption{Category-dependent samples per training geometry. Airplane condition groups share the same four-cloud, ten-condition sampling scheme.}
\label{tab:categorysampling}
\begin{tabular}{lrrrp{0.47\linewidth}}
\toprule
\textbf{Category} & \textbf{Clouds} & \textbf{Conditions/cloud} & \textbf{Samples} & \textbf{Inflow and condition range}\\
\midrule
Car & 40 & 1 & 40 & Fixed $+x$ inflow.\\
Airplane & 4 & 10 & 40 & NASA-like: angle of attack $[-2.5^\circ,5^\circ]$, Mach code $[0.5,0.9]$. AirCraft-like: Mach code $[2,7]$, nonzero angle of attack $[0^\circ,7^\circ]$ and sideslip $[0^\circ,2^\circ]$; also includes fixed-Mach and zero-angle cases.\\
Watercraft & 4 & 10 & 40 & Yaw sampled uniformly from $[-10^\circ,10^\circ]$.\\
\bottomrule
\end{tabular}
\end{table}
\FloatBarrier

The corpus contains 8,971 training geometries with 40 samples each, giving 358,840 training examples. The held-out set contains 100 geometries per category and two validation examples per geometry. Given our computational budget, we reproduce GeoPT's pre-training method on this shared corpus and training schedule rather than replicate the full data scale and experimental scope of the original study (\cref{app:trainingconfig}). Cars vary the query cloud rather than the inflow; airplanes and watercraft reuse each cloud across conditions. Mach codes do not introduce compressible physics into the potential-flow labels. Input encodings and losses are specified in \cref{app:interface}.

\subsection{Labeling Cost}
\label{app:cost}
\label{app:costcomparison}
The timing unit is one geometry--condition--query-cloud example. We measure four clouds with ten conditions each on one car, airplane, and watercraft, for 120 examples. Reported per-example times are amortized over this benchmark layout. Neural training is excluded.

\paragraph{Work and reuse.} With $P$ panels, dense assembly costs $O(P^2)$ and factorization costs $O(P^3)$. Three inflow bases share the factorization. Surface and volume basis evaluations cost $O(P^2)$ and $O(Q_vP)$, respectively; each additional condition costs $O(Q_v+Q_s)$ to combine velocities and form $C_p$. GeoPT reuses geometry acceleration structures and initial VDFs, but each motion requires collision checks and updated distance queries. LBM advances $N$ lattice cells for $T$ steps, with work proportional to $NT$; a solved field can serve additional query clouds.

\paragraph{Matched CPU measurements.} We measure labeling from prepared meshes and query clouds on one geometry per category (\cref{tab:costgeometries}). SimpFlow and our GeoPT reproduction use the same query clouds, each with 8,192 volume and 4,096 surface points, and the same four logical CPUs of an AMD EPYC 9684X. For each geometry, the elapsed time covers generation of all 40 examples, including method-specific initialization and reuse across examples.

The timing benchmark uses four clouds and ten conditions or motions per cloud for both methods. SimpFlow uses its training condition samplers for airplanes and watercraft, and varying yaw for the car timing test. This common timing layout differs from the car training corpus, which uses 40 clouds at fixed inflow.

\begin{table}[!ht]
\centering\appendixtablestyle
\caption{Labeling costs and hardware. SimpFlow (4k) and our GeoPT reproduction use four logical CPUs of an AMD EPYC 9684X; category rows report seconds per 40-example batch. LBM is a literature-based estimate on one A100, reported only as amortized time per example.}
\label{tab:costgeometries}
\begin{tabular}{llrrr}
\toprule
\textbf{Category} & \textbf{ShapeNet ID} & \textbf{SimpFlow} & \textbf{GeoPT} & \textbf{LBM}\\
\midrule
Car & \texttt{7e7f0637d6798113ddb4014f0e8d0dac} & 5.195 & 21.655 & ---\\
Airplane & \texttt{870dc1667e957672c66e7238ddb322f4} & 4.032 & 6.831 & ---\\
Watercraft & \texttt{8411cba605709e46c26de8baaf779d69} & 4.050 & 8.571 & ---\\
\midrule
\multicolumn{2}{l}{Total for 120 examples (s)} & 13.277 & 37.057 & ---\\
\multicolumn{2}{l}{Amortized time (s/example)} & \textbf{0.111} & 0.309 & 135.5$^{*}$\\
\multicolumn{2}{l}{Hardware} & 4 CPU threads & 4 CPU threads & 1 A100\\
\bottomrule
\end{tabular}
\par\smallskip
\begin{minipage}{.96\linewidth}\small
$^{*}$From the reported LBM solve time~\citep{shen2026scaling3d}, assuming four query clouds per solution: $542/4=135.5$\,s/example. Dashes indicate no measurement on the listed geometries.
\end{minipage}
\end{table}

\paragraph{Amortization.} For method $m$ and geometry $g$, let $T_{m,g}$ be the measured 40-example labeling time. The amortized time per example is
\begin{equation}
\overline t_m=\frac{\sum_{g=1}^{3}T_{m,g}}{3\times4\times10}.
\end{equation}
SimpFlow totals 13.277\,s, or 0.111\,s per example; GeoPT totals 37.057\,s, or 0.309\,s. The ratio of unrounded means is 2.79. Per-geometry times for SimpFlow are 0.130, 0.101, and 0.101\,s, compared with 0.541, 0.171, and 0.214\,s for GeoPT. Under the same protocol, 2k and 8k cost 0.055 and 0.322\,s per example. Phase totals in \cref{tab:labelcost} show that source solving and surface bases dominate the optimized 4k pipeline.

\begin{table}[!ht]
\centering\appendixtablestyle
\caption{Label-generation time by stage, summed over the three geometries. Each method's total is divided by 120 to obtain the per-example time in \cref{tab:threewaycost}.}
\label{tab:labelcost}
\begin{tabular}{llr}
\toprule
\textbf{Method} & \textbf{Phase} & \textbf{Total time (s)}\\
\midrule
SimpFlow & Geometry solves and panel-surface velocity bases & 10.279\\
& Volume-field evaluation for 12 query clouds & 2.876\\
& Condition generation and 120 label combinations & 0.121\\
\midrule
GeoPT & Geometry acceleration structures & 0.114\\
& Initial distance fields for 12 query clouds & 0.324\\
& Motion-label generation for 120 examples & 36.619\\
\bottomrule
\end{tabular}
\end{table}

\paragraph{LBM estimate.} \citet{shen2026scaling3d} report 542\,s on one A100 for a $300\times120\times90$ lattice and 16,000 steps, producing surface pressure and wall shear. For $K$ conditions distributed over 40 sampled examples, solve-only amortization is $542K/40$\,s plus label extraction. Table~\ref{tab:threewaycost} assumes four clouds share each solution, giving $542\times10/40=135.5$\,s per example, with extraction omitted. Under this reuse assumption, the estimate is about 1,225 times SimpFlow's measured 4k mean on the stated GPU/CPU hardware. The comparison concerns labeling time for different source models and output fields; it does not establish a transfer-accuracy-matched speedup. SimpFlow's basis reuse also avoids a new flow solve when the inflow direction changes.
\FloatBarrier

\FloatBarrier
\section{Experimental Setup}
\label{app:experiments}
\subsection{Data Splits and Evaluation}
\begin{table}[ht]
\centering\appendixtablestyle
\caption{Five flow tasks and a separate structural-response test, with nested training budgets and fixed test sets. Counts refer to complete simulation cases, not query points.}
\label{tab:splits}
\begin{tabular}{llrrr}
\toprule
\textbf{Benchmark} & \textbf{Prediction target} & \textbf{Train budgets} & \textbf{Validation} & \textbf{Test}\\
\midrule
AirCraft & Surface $C_p$ & 20--100 & 25 & 25 \\
NASA-CRM & Surface $C_p$ & 20--100 & 5 & 44 \\
DrivAerML surface & Surface $C_p$ & 20--100 & 25 & 25 \\
DTCHull & Volume $\mathbf v$ & 20--100 & 9 & 10 \\
DrivAerML volume & Volume $[C_p,\mathbf v]$ & 10, 20, 30, 40 & 0 & 7 \\
\midrule
\multicolumn{5}{l}{Separate structural-response test}\\
Car-Crash & Surface $\sigma_{\mathrm{VM}}^{\max}$ & 20, 40, 60, 80 & 17 & 34 \\
\bottomrule
\end{tabular}
\end{table}
AirCraft uses 30 geometries in ten families of three, with five conditions per geometry. Two geometries per family enter training; the remaining geometries are split across validation and test. Training budgets balance geometry families and conditions. NASA-CRM retains its official 44-case test set; five of the 105 available training cases form validation, and the remaining 100 are ordered by farthest-point sampling in standardized Mach--angle-of-attack space. DrivAerML and DTCHull use fixed random geometry-disjoint splits. Every compared method uses the same split and nested budgets. The maximum budget is the selected training pool, not necessarily the full source dataset.

\paragraph{Downstream preprocessing.}
Each of the original four CFD tasks contains one stored set of 4,096 points (\cref{tab:downstreampoints}). Points are sampled uniformly without replacement from the native vertices or cells, rather than by surface area. AirCraft reuses the same vertex indices across the five conditions of a geometry. DTCHull volume candidates lie in the model-coordinate box $[-4,5]\times[-1,1]\times[-1.5,1.5]$; its 512 surface points provide geometry context but receive no downstream loss. All methods load the same arrays throughout training and evaluation, with no online point resampling or coordinate augmentation.

\begin{table}[ht]
\centering\appendixtablestyle

\caption{Fixed downstream point sets and dimensionless flow prediction targets.}
\label{tab:downstreampoints}
\begin{tabular}{lp{0.38\linewidth}p{0.36\linewidth}}
\toprule
\textbf{Benchmark} & \textbf{Point set} & \textbf{Prediction target}\\
\midrule
AirCraft & 4,096 native surface vertices & Dataset-provided $C_p$ \\
NASA-CRM & 4,096 native surface vertices & Dataset-provided $C_p$ \\
DrivAerML & 4,096 native surface-cell centers & Surface pressure coefficient $C_p$ \\
DTCHull & 3,584 native volume points and 512 surface vertices & Normalized volume velocity $\mathbf v$ \\
\bottomrule
\end{tabular}
\end{table}

Geometry coordinates are centered at the bounding-box midpoint and scaled isotropically to streamwise length five; CFD queries use the same transform. NASA-CRM uses the axis ordering $(x,z,y)$, and DTCHull uses $(-x,z,y)$, with the latter also applied to velocity components. The distance and direction channels are computed in these model coordinates, and the inflow encodings are given in \cref{app:interface}. Only the targets listed in \cref{tab:downstreampoints} enter the loss and test metric; other output components are masked in these four tasks.

Native pressure $p$ and velocity $\mathbf u$ are nondimensionalized using \cref{eq:nondimensional}: pressure becomes $C_p=(p-p_\infty)/(\tfrac12\rho U_\infty^2)$ and velocity becomes $\mathbf v=\mathbf u/U_\infty$, with case-specific freestream reference values. Dataset-provided pressure coefficients are used directly. The task tables list these dimensionless flow targets. Pretraining and flow fine-tuning therefore use the same output quantities, with task-specific observation masks.

\paragraph{Additional tasks.} DrivAerML volume uses a frozen 47-case pool: 40 training cases and seven test cases, without a validation set. Budgets 10, 20, 30, and 40 use nested prefixes of the frozen training list. Each case contains 8,192 volume queries and 4,096 surface context points, with no extra wake-focused sampling. Only volume points enter the loss, and all four quantities $[C_p,\mathbf v]$ are supervised without component masking. Its reported joint relative $L_2$ uses this dimensionless four-channel vector; surface context points receive no loss. Model selection uses training-set four-channel MSE.

Car-Crash uses the undeformed surface geometry, 4,096 surface queries, and scalar maximum-over-time von Mises stress. Its original 80/17/34 train/validation/test partition is fixed, and training budgets are the first 20/40/60/80 entries of the frozen training list. Model selection uses validation relative $L_2$. The existing 11-channel inputs are retained for both added tasks. Neither additional task uses test data for fitting or model selection.

For one test case and its valid observed components $\mathcal O$, the reported error is
\begin{equation}
e=\frac{\left(\sum_{(i,k)\in\mathcal O}(\widehat y_{i,k}-y_{i,k})^2\right)^{1/2}}
{\left(\sum_{(i,k)\in\mathcal O}y_{i,k}^2\right)^{1/2}}.
\end{equation}
We average $e$ over complete test cases. Every main comparison uses fine-tuning seed 42 and, when applicable, pre-training seed 42. We report no across-seed standard deviation or significance test. Main tables use fixed epoch 200; model-selection checkpoints are not substituted.

\subsection{Prediction Interface}
\label{app:interface}
For a volume query, let $\mathbf r_i$ point from $\mathbf x_i$ to its nearest surface point. We use $d_i=\|\mathbf r_i\|$ and $\boldsymbol\eta_i=\mathbf r_i/d_i$. At surface points, $d_i=0$ and $\boldsymbol\eta_i$ is the face normal. The volume convention is not an outward-normal field. Reversing $\boldsymbol\eta_i$ leaves the local geometry projection $(\mathbf a\cdot\boldsymbol\eta_i)\boldsymbol\eta_i$ unchanged, but the direction supplied to the network follows the stated convention.

The condition channels follow the car and aerodynamics encodings in GeoPT's implementation~\citep{wu2026geopt}. With angle of attack $\alpha$, sideslip $\beta$, yaw $\psi$, and Mach code $M$, the direction and scalar encodings are listed in \cref{tab:conditions}.
\begin{table}[ht]
\centering\appendixtablestyle
\caption{Condition channels. Directions use the corresponding benchmark coordinate frame; angles in these expressions are in radians.}
\label{tab:conditions}
\begin{tabular}{lll}
\toprule
\textbf{Task} & \textbf{Unit direction $\mathbf a$} & \textbf{Scalar $a_c$}\\
\midrule
Car & $(1,0,0)$ & $0.3$\\
NASA-CRM & $(\cos\alpha,\sin\alpha,0)$ & $1.6M$\\
AirCraft & $(\cos\alpha\cos\beta,\sin\alpha,\cos\alpha\sin\beta)$ & $M/3$\\
Watercraft pre-training & $(\cos\psi,0,\sin\psi)$ & $0.3$ at all queries\\
Hull fine-tuning & $(\cos\psi,0,\sin\psi)$ & $0.3$ in water, $0$ in air\\
\bottomrule
\end{tabular}
\end{table}
\FloatBarrier
Inflow direction is broadcast across queries. The scalar is also global in car and aircraft tasks, but hull fine-tuning uses the water/air region encoding supplied with the preprocessed data. Upstream watercraft do not use this air-side gate. Thus, the attribution concerns the encoding conventions and empirical direction/scale guidance, not an identical pre-training strategy. The potential teacher uses unit inflow and does not model Mach- or Reynolds-dependent effects.

Both stages use the masked MSE in \cref{eq:loss}, with $\mathcal O_{\mathcal B}$ containing valid sample--query--component indices in batch $\mathcal B$. Invalid low-fidelity labels and missing downstream observations are excluded componentwise; geometry-context points without target observations receive no loss. The query set is $\mathcal Z_b=\{\mathbf z_i\}$. Flow fine-tuning inherits and updates all parameters, while Car-Crash inherits the embedding and backbone and initializes a new scalar output head.

\FloatBarrier
\subsection{Training Configuration}
\label{app:trainingconfig}
All methods share the backbone and optimization settings in \cref{tab:trainingconfig}. The same 4k initialization is reused across all downstream tasks and budgets.

\begin{table}[!ht]
\centering\appendixtablestyle
\caption{Shared model and training configuration. All main comparisons use seed 42. Each pre-training epoch traverses all 40 samples per geometry.}
\label{tab:trainingconfig}
\begin{tabular}{p{.29\linewidth}p{.31\linewidth}p{.31\linewidth}}
\toprule
\textbf{Setting} & \textbf{Pre-training} & \textbf{Fine-tuning}\\
\midrule
Backbone & \multicolumn{2}{l}{Transolver: 8 layers, width 256, 8 heads}\\
Slice tokens & \multicolumn{2}{l}{32 per head}\\
Epochs & 8 & 200\\
Optimizer updates & 358,840 & Budget-dependent\\
Batch size & 8 & 1\\
Optimizer & AdamW & AdamW\\
Learning rate & $3\times10^{-4}$ & Peak $10^{-3}$\\
Schedule & Warmup + cosine & OneCycle\\
Weight decay & $10^{-5}$ & $10^{-5}$\\
Loss & MSE & MSE\\
\midrule
GPU & \multicolumn{2}{l}{NVIDIA RTX 6000 Ada (48\,GB)}\\
GPUs per run & 1 & 1\\
Approx. GPU-hours (SimpFlow, 4k) & $\approx25$ & 0.026--0.135\\
\bottomrule
\end{tabular}
\end{table}

Runtime entries refer to the 4k model. Pre-training time combines the logged training segments, including resumed work but excluding server downtime. Fine-tuning time is the range per 200-epoch run across all six benchmarks and their budgets (seed 42), including evaluation and checkpointing. Each run uses one GPU; independent runs may execute concurrently. Label-generation costs are reported separately in \cref{app:cost}.

\FloatBarrier
\subsection{GeoPT Main Baseline}
\label{app:reproductionscope}
Resource constraints require a smaller pre-training corpus and fewer query points than in the original GeoPT study~\citep{wu2026geopt}. We retrain GeoPT under the same data scale as LGF and SimpFlow (\cref{tab:reproductionscope}); all main downstream comparisons, including scratch, share the same splits, query points, backbone, and optimization settings.

\begin{table}[!ht]
\centering\appendixtablestyle
\caption{Actual configurations used in our main experiments. Shared optimization settings are listed in \cref{tab:trainingconfig}.}
\label{tab:reproductionscope}
\begin{tabular}{p{.35\linewidth}p{.25\linewidth}p{.30\linewidth}}
\toprule
\textbf{Setting} & \textbf{GeoPT (retrained)} & \textbf{LGF / SimpFlow}\\
\midrule
Pre-training geometries & 8,971 & 8,971\\
Stored training samples & 358,840 & 358,840\\
Pre-training volume / surface points & 8,192 / 4,096 & 8,192 / 4,096\\
Layers / width & 8 / 256 & 8 / 256\\
Downstream points & \multicolumn{2}{l}{4,096; 12,288 for DrivAerML volume}\\
\bottomrule
\end{tabular}
\end{table}
\FloatBarrier

GeoPT outperforms scratch in all 24 main task--budget settings at epoch 200 (\cref{tab:fullresults}). The configuration above applies to the main comparisons with LGF and SimpFlow. The earlier reproduction and input-head control use the separate protocol below.

\FloatBarrier
\Needspace{.65\textheight}
\subsection{GeoPT Reproduction}
\label{app:earlygeopt}
\begin{figure}[!ht]
\centering
\includegraphics[width=.92\linewidth]{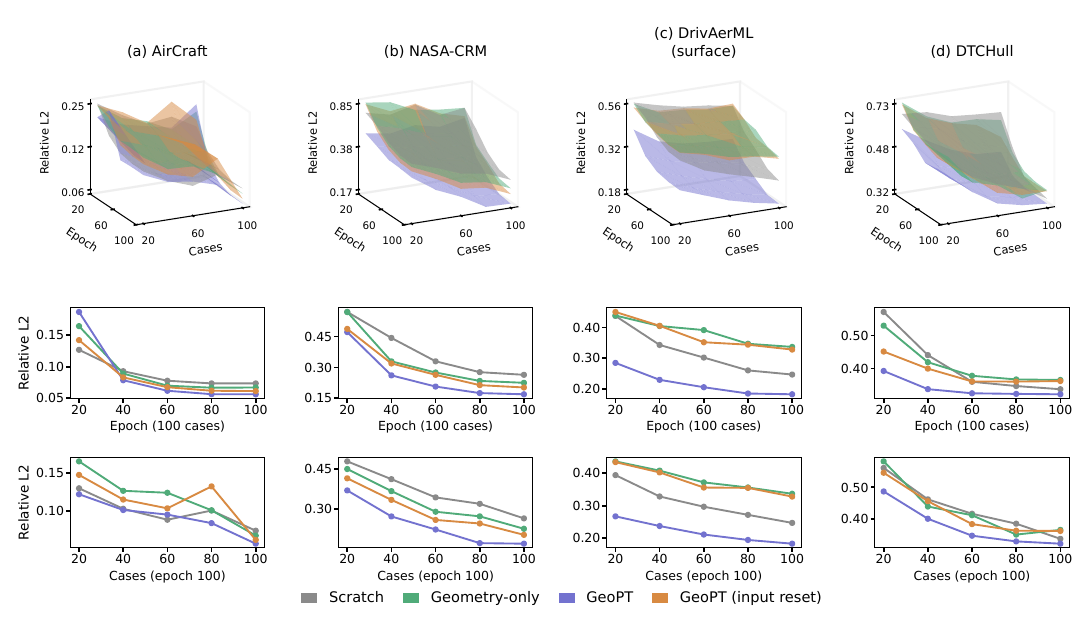}
\caption{Early GeoPT reproduction and input-head control under one matched protocol. Top: test relative $L_2$ across budgets and epochs (logarithmic height). Middle: epoch curves at 100 cases. Bottom: budget curves at epoch 100. Pre-training uses 4,096 volume and 512 surface points; fine-tuning uses 4,096 points. Original joint targets have 6/1/1/4 channels for AirCraft/NASA-CRM/DrivAerML/DTCHull. The reset variant retains GeoPT's backbone and reinitializes its input mapping. All reported counts use epoch-100 test errors.}
\label{fig:geoptpilot}
\end{figure}

An independent early study uses an 8-layer, 128-channel model, 448 unlabeled downstream geometries, 50 pre-training epochs, and 100 fine-tuning epochs (seed 42), separately from the main protocol. It recovers GeoPT's core qualitative findings: geometry-only pre-training can cause negative transfer (11/20 settings), whereas GeoPT improves over scratch in 19/20 (\cref{fig:geoptpilot}). Keeping the same pre-trained backbone but resetting the input mapping raises test error in all 20 settings and produces negative transfer in 10/20. Both GeoPT variants reset the output head; only input-mapping inheritance differs. This shows the importance of transferring the learned input mapping together with the backbone, as we do in the main GeoPT baseline. The reduced experimental scale reflects resource constraints.

\FloatBarrier

\FloatBarrier
\Needspace{.90\textheight}
\section{Flow Structure and Label Accuracy}
\label{app:fieldanalysis}
We first analyze how potential flow and LGF differ in velocity and pressure structure, then compare fields on actual geometries and quantify their label errors against CFD (\cref{sec:why}).

\subsection{Velocity and Pressure Structure}
\label{app:smooth}
The LGF baseline uses $\delta=0.5$ in normalized coordinates. At smooth locations with a unique nearest point, let $d$ be the exterior distance and $\mathbf n_d=\nabla d$ with unit norm. The volume-point convention gives $\boldsymbol\eta=-\mathbf n_d$, which leaves the projection unchanged. At $d=0$, $\omega=1$ and the ideal LGF has zero normal velocity. Wall impermeability alone therefore does not distinguish this baseline from potential flow.
Writing $t=\mathbf a\cdot\mathbf n_d$ gives
\begin{equation}
\|\mathbf v_s\|^2=1-2\omega t^2+\omega^2t^2,
\qquad C_{p,s}=\omega(2-\omega)t^2\in[0,1].
\end{equation}
This baseline attenuates the inflow but cannot accelerate it above unit magnitude. Its targets do not activate the pressure or speed clipping limits.
Let $H_d$ denote the distance Hessian. Using $H_d\nabla d=0$, differentiation yields
\begin{align}
\nabla\cdot\mathbf v_s&=-[\omega'(d)+\omega(d)\Delta d](\mathbf a\cdot\nabla d),\\
\nabla\times\mathbf v_s&=-\omega(d)(H_d\mathbf a)\times\nabla d.
\end{align}
These quantities are generally nonzero. These derivatives apply where the distance is differentiable. On a sphere, ideal potential flow gives surface $C_{p,L}=1-\tfrac94\sin^2\vartheta$, whereas this local geometry rule gives $C_{p,s}=\cos^2\vartheta$. At the equator they equal $-1.25$ and $0$, respectively, illustrating global flow acceleration versus local attenuation.

\subsection{Reference Fields}
All potential-flow illustrations use their original 8k solutions. Shared color limits reveal stagnation, suction, and acceleration patterns; the annotated errors are computed before display averaging and color clipping. Car and aircraft annotations use the same full surface query set for both labels. Detailed metric definitions and diagnostic exclusions are in \cref{app:referenceaccuracy}.

\begin{figure}[!ht]
\centering
\includegraphics[width=.70\textwidth]{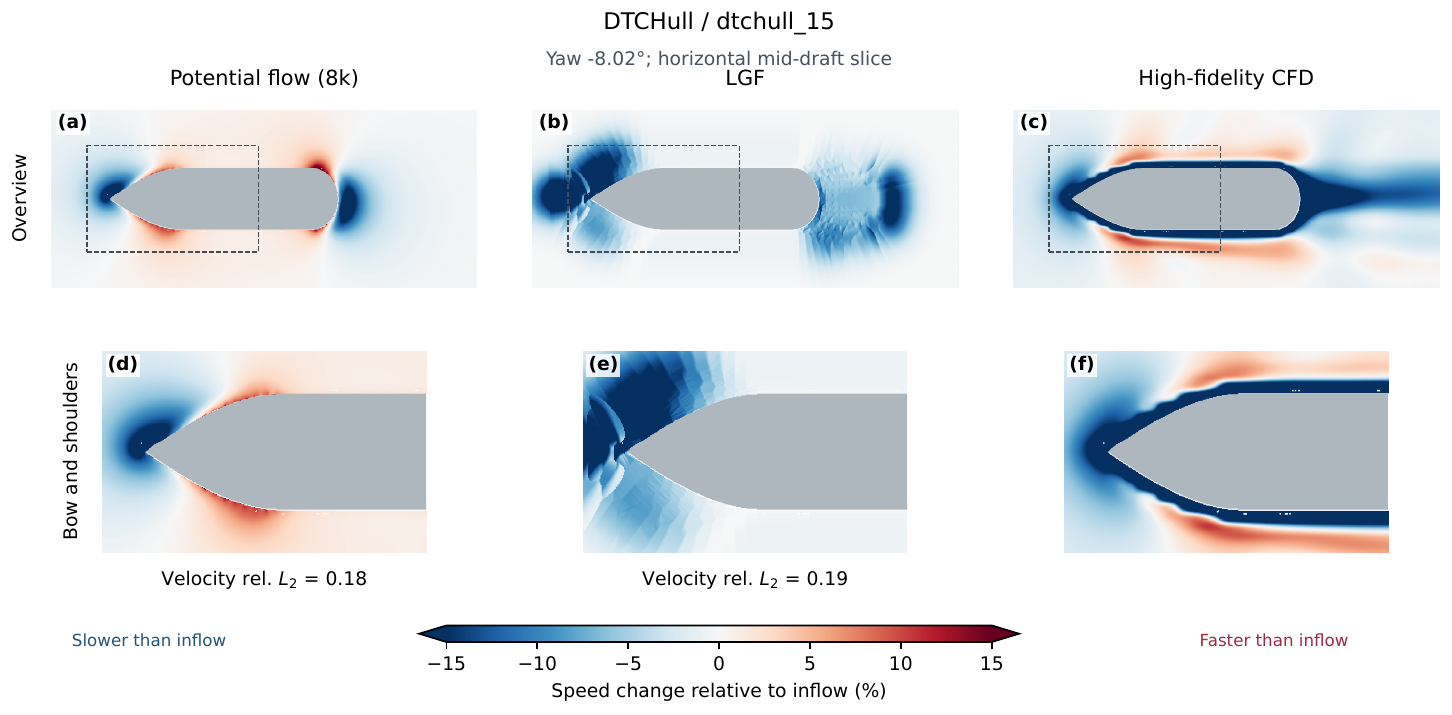}
\caption{DTCHull \texttt{dtchull\_15}, 8k double-body potential flow, LGF, and CFD at matched yaw $-8.02^\circ$ and $U_\infty=1.668$. The inlet phase interface sets the waterline; the mid-draft slice is at $y=-0.1293$. Rows show the full slice and bow detail. Colors are $100(|\mathbf u|/U_\infty-1)$, clipped to $[-15\%,15\%]$; gray marks the hull. The $901\times373$ display grid uses direct potential evaluation and native CFD interpolation. CFD colors are lightly smoothed for display with a Gaussian filter ($\sigma=3$ grid points); error metrics use unsmoothed fields. Annotated vector errors use the original 31,696 common queries before clipping, separately from formal test-set metrics.}
\label{fig:hullreference}
\end{figure}

On DTCHull, potential flow and CFD both distinguish bow deceleration from shoulder acceleration, whereas LGF only reduces the inflow speed and misses the accelerated regions (\cref{fig:hullreference}). This spatial redistribution supplies a useful flow prior for hull transfer. However, potential flow imposes impermeability with tangential slip: it does not reproduce the viscous near-wall low-speed layer or the extended downstream velocity deficit visible in CFD. The double-body approximation also omits free-surface dynamics.
\FloatBarrier
\begin{figure*}[p]
\centering
\includegraphics[width=.575\textwidth]{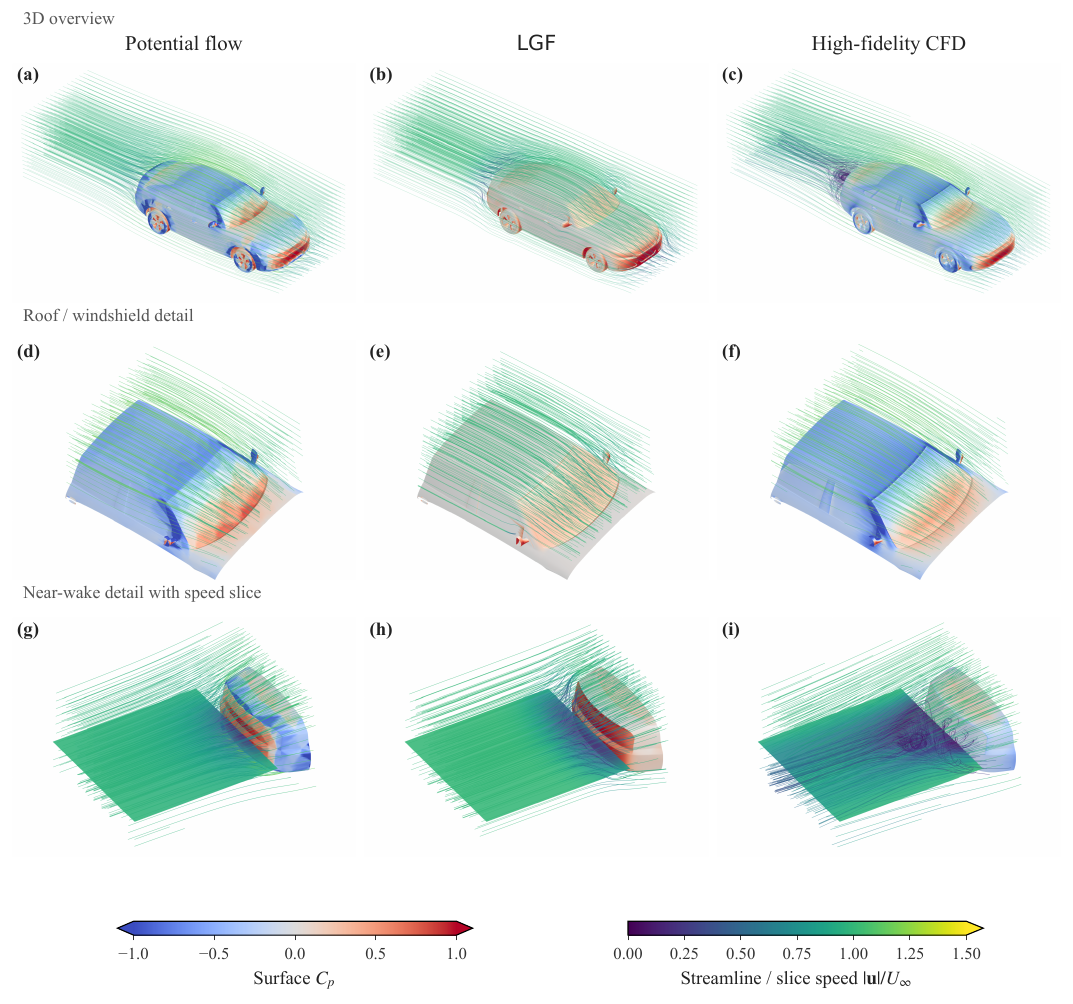}
\caption{Detailed reference-field comparison for the same car and inflow. Columns show potential flow, LGF, and high-fidelity CFD. Rows show an overview, roof/windshield details, and near-wake details with a speed slice. Surface $C_p$ uses the shared range $[-1,1]$; normalized speed uses $[0,1.5]$. Cameras, seeds, and scales are matched. Surface rendering uses the full boundary reference, while streamline integration uses reconstructed steady or mean velocity fields. The potential solution retains selected large-scale flow features despite large pointwise discrepancies from CFD; visual similarity is not quantitative agreement.}
\label{fig:physicaldetails}

\medskip
\includegraphics[width=.485\textwidth]{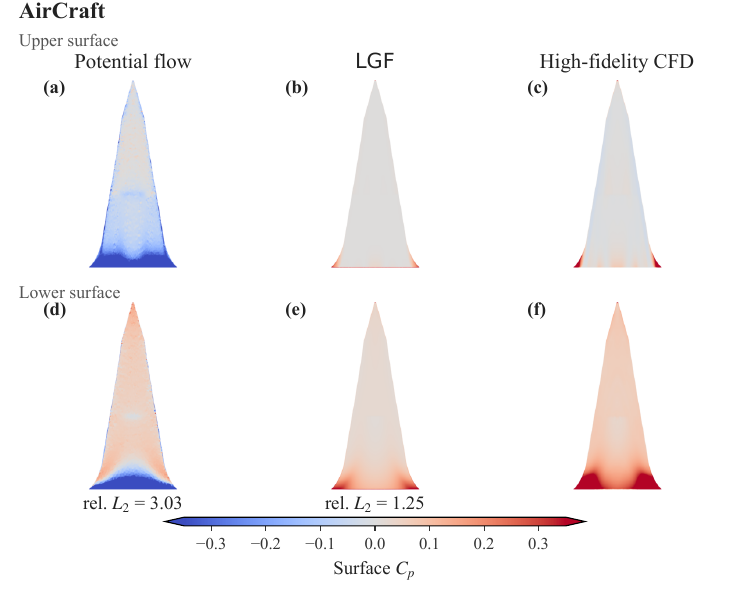}\hfill
\includegraphics[width=.485\textwidth]{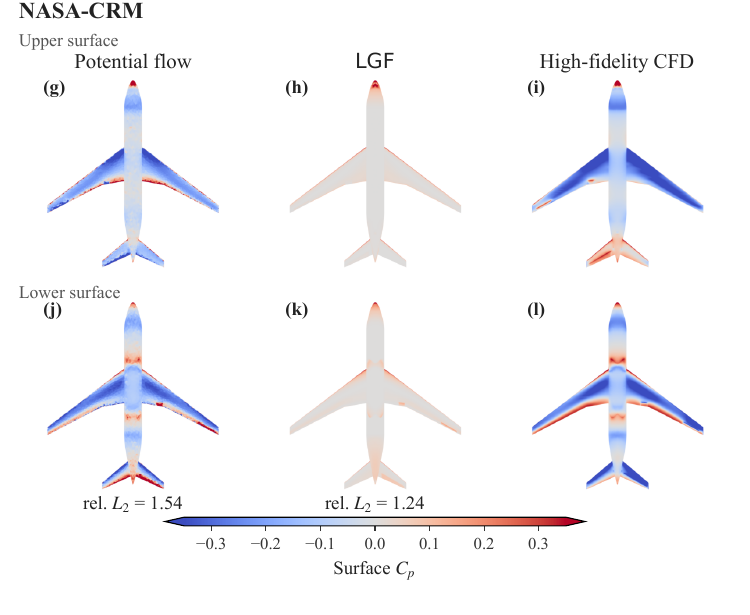}
\caption{Aircraft surface references: AirCraft aircraft\_22 ((a)--(f); Mach 7, angle of attack $7^\circ$, zero sideslip) and NASA-CRM train\_Sample001 ((g)--(l); Mach 0.7, zero angle of attack). Columns: 8,000-panel potential flow, LGF, and CFD; rows: upper/lower surfaces. The $C_p$ color scale is clipped to $[-0.35,0.35]$ for all panels. Potential flow produces both high-pressure and negative-pressure suction regions, whereas LGF has $C_p\geq0$ and cannot represent suction. On NASA-CRM, potential flow captures the nose high pressure and wing suction seen in CFD; LGF retains nose high pressure but misses wing suction. Potential flow nevertheless differs from CFD in tail pressure. On AirCraft, potential flow distinguishes upper-surface suction from lower-surface compression, but predicts negative rear pressure where CFD shows positive pressure. It therefore retains part of the pressure organization without reproducing every region or its magnitude. Color clipping hides amplitude differences; the shared scale is used to compare spatial patterns. Annotated relative $L_2$ values show that pointwise errors remain large. For each sample, both labels use the same full set of surface queries without label-dependent exclusions; each column shows its whole-surface value once below the lower-surface row. Metrics precede face-to-vertex display averaging, with lighting disabled. LGF labels use $d=0$ and original face normals. Full per-dataset and per-sample errors are in \cref{app:referenceaccuracy}.}
\label{fig:aircraftreference}
\label{fig:nasareference}
\label{app:aircraftreference}
\end{figure*}
\FloatBarrier
\subsection{Reference-Label Accuracy}
\label{app:referenceaccuracy}
\begingroup
\small
\setlength{\intextsep}{5pt}
\setlength{\textfloatsep}{5pt}
Figure metrics use the exact dense face-centroid queries before display averaging and clipping. Potential labels retain their original $[-2,2]$ bound; CFD is unmodified. Relative $L_2$ and MAE use all finite points. Figure annotations use this full set; the diagnostic ``excluded'' column removes points with the evaluated label's $C_p\leq-2$ and can therefore use different subsets across labels.
\begin{table}[!ht]
\centering\appendixtablestyle
\caption{Surface $C_p$ errors for the exact fidelity-figure samples. AirCraft is 22\_7.0\_7.0\_0.0 (663,938 points), NASA-CRM is train\_Sample001 (908,804), and DrivAerML is run\_181 (753,246). ``Excluded'' removes label $C_p\leq-2$; MAE uses all points. The figures display LGF and 8k potential labels; 2k/4k rows use other existing labels for the same queries.}
\label{tab:figurelabelerrors}
\begin{tabular}{llrrrrr}
\toprule
\textbf{Dataset} & \textbf{Label} & \textbf{$L_2$ all} & \textbf{$L_2$ excluded} & \textbf{Removed} & \textbf{Removed (\%)} & \textbf{MAE}\\
\midrule
DrivAerML & LGF & 1.5623 & 1.5623 & 0 & 0.000 & 0.5743\\
DrivAerML & Potential flow (2k panels) & 1.7037 & 1.7851 & 35446 & 4.706 & 0.6265\\
DrivAerML & Potential flow (4k panels) & 1.6628 & 1.8266 & 51658 & 6.858 & 0.6186\\
DrivAerML & Potential flow (8k panels) & 1.6114 & 1.7961 & 34644 & 4.599 & 0.6063\\
\midrule
AirCraft & LGF & 1.2495 & 1.2495 & 0 & 0.000 & 0.0618\\
AirCraft & Potential flow (2k panels) & 3.0961 & 2.5703 & 11871 & 1.788 & 0.1951\\
AirCraft & Potential flow (4k panels) & 3.0861 & 2.5206 & 12648 & 1.905 & 0.2012\\
AirCraft & Potential flow (8k panels) & 3.0299 & 2.5524 & 10871 & 1.637 & 0.2010\\
\midrule
NASA-CRM & LGF & 1.2412 & 1.2412 & 0 & 0.000 & 0.2651\\
NASA-CRM & Potential flow (2k panels) & 1.2962 & 1.2125 & 3908 & 0.430 & 0.2509\\
NASA-CRM & Potential flow (4k panels) & 1.4477 & 1.1985 & 17149 & 1.887 & 0.2845\\
NASA-CRM & Potential flow (8k panels) & 1.5400 & 1.4223 & 6948 & 0.765 & 0.3087\\
\bottomrule
\end{tabular}
\end{table}

The car's volume errors use velocity vectors at native CFD source nodes before streamline interpolation. The 2k/4k rows retain their saved exterior-valid masks, hence their different point counts. Aircraft figures show pressure only.
\begin{table}[!ht]
\centering\appendixtablestyle
\caption{DrivAerML run\_181 volume-vector errors associated with the car reference visualization. The figure uses LGF and 8k labels; other rows are existing resolution-study labels.}
\label{tab:figurevelocityerrors}
\begin{tabular}{lrrrrr}
\toprule
\textbf{Label} & \textbf{Valid points} & \textbf{$L_2$ all} & \textbf{$L_2$ excluded} & \textbf{Removed} & \textbf{Removed (\%)}\\
\midrule
LGF & 62652 & 0.4500 & 0.4500 & 0 & 0.000\\
Potential flow (2k panels) & 61331 & 0.4808 & 0.4692 & 229 & 0.373\\
Potential flow (4k panels) & 61955 & 0.4939 & 0.4727 & 418 & 0.675\\
Potential flow (8k panels) & 62652 & 0.4819 & 0.4723 & 232 & 0.370\\
\bottomrule
\end{tabular}
\end{table}

\paragraph{Full test-set label errors.}
Table~\ref{tab:datasetlabelerrors} uses the fine-tuning test splits, sampled queries, inflows, and normalization. Entries are case means and population standard deviations. The no-wake proxy removes the last 15\% of inflow-projected geometry extent for surface tasks, and points downstream of the stern for DTCHull. These geometric masks retain clipped labels and differ from the dense figure queries.
\begin{table}[!ht]
\centering\appendixtablestyle
\caption{Existing full-test-set label relative $L_2$ (case mean $\pm$ population standard deviation). Numbers of cases: DrivAerML 25, AirCraft 25, NASA-CRM 44, DTCHull 10. A dash denotes an unavailable strict full-case statistic, not zero error.}
\label{tab:datasetlabelerrors}
\begin{tabular}{llrrrr}
\toprule
\textbf{Dataset} & \textbf{Region} & \textbf{LGF} & \textbf{\shortstack{Potential flow\\(2k)}} & \textbf{\shortstack{Potential flow\\(4k)}} & \textbf{\shortstack{Potential flow\\(8k)}}\\
\midrule
DrivAerML & Full & $1.4251\pm0.0360$ & $1.4024\pm0.0669$ & $1.4090\pm0.0669$ & $1.3837\pm0.0617$\\
 & No-wake proxy & $1.3887\pm0.0313$ & $1.3422\pm0.0643$ & $1.3468\pm0.0713$ & $1.3207\pm0.0595$\\
\midrule
AirCraft & Full & $1.9734\pm0.2265$ & $3.9812\pm0.6189$ & $4.0537\pm0.6907$ & $4.1088\pm0.7196$\\
 & No-wake proxy & $0.5498\pm0.0798$ & $1.8876\pm0.5995$ & $2.1291\pm0.8539$ & $2.1597\pm0.9022$\\
\midrule
NASA-CRM & Full & $1.2388\pm0.0708$ & $1.2890\pm0.1991$ & $1.4618\pm0.2551$ & $1.6227\pm0.2601$\\
 & No-wake proxy & $1.2441\pm0.0810$ & $1.2751\pm0.2115$ & $1.5089\pm0.2679$ & $1.7124\pm0.3069$\\
\midrule
DTCHull & Full & $0.1801\pm0.0284$ & --- & --- & ---\\
 & No-wake proxy & $0.1981\pm0.0318$ & --- & --- & ---\\
\bottomrule
\end{tabular}
\end{table}

DTCHull potential-flow entries are unavailable under the strict full-case criterion: some CFD queries lie inside simplified meshes, and six of ten original closed hulls have fewer than 8,000 faces. Missing labels are not filled. Numerical refinement of the same physical model does not remove its modeling error relative to CFD.
\FloatBarrier
\endgroup

\FloatBarrier
\section{Flow Transfer Results}
\label{app:additionalresults}
\subsection{Final Errors}
\label{app:learninggrid}
\Cref{tab:fullresults} gives all 24 flow task--budget comparisons at epoch 200. \Cref{fig:transfer} shows how test error varies with training budget and fine-tuning duration.
\begin{table}[!ht]\centering\appendixtablestyle
\caption{Test relative $L_2$ at epoch 200, using pre-training and fine-tuning seed 42. Values are means over test cases, not over seeds. SimpFlow uses 4k panels. Bold marks the lowest error in each row.}
\label{tab:fullresults}
\begin{tabular}{llrrrr}\toprule
\textbf{Benchmark} & \textbf{Cases} & \textbf{Scratch} & \textbf{LGF} & \textbf{GeoPT} & \textbf{SimpFlow}\\\midrule
AirCraft & 20 & 0.2254 & 0.2174 & 0.2063 & \textbf{0.1838}\\
AirCraft & 40 & 0.2128 & 0.1855 & 0.1685 & \textbf{0.1655}\\
AirCraft & 60 & 0.2409 & 0.1611 & 0.1571 & \textbf{0.1541}\\
AirCraft & 80 & 0.2200 & 0.1633 & \textbf{0.1489} & 0.1492\\
AirCraft & 100 & 0.1587 & 0.1501 & \textbf{0.1335} & 0.1468\\
\midrule
NASA-CRM & 20 & 0.3919 & 0.3111 & 0.2783 & \textbf{0.2213}\\
NASA-CRM & 40 & 0.2617 & 0.2118 & 0.1697 & \textbf{0.1592}\\
NASA-CRM & 60 & 0.1881 & 0.1769 & 0.1622 & \textbf{0.1288}\\
NASA-CRM & 80 & 0.1422 & 0.1481 & 0.1169 & \textbf{0.1160}\\
NASA-CRM & 100 & 0.1258 & 0.1295 & \textbf{0.1042} & 0.1054\\
\midrule
DrivAerML (surface) & 20 & 0.2863 & 0.3024 & 0.2214 & \textbf{0.2128}\\
DrivAerML (surface) & 40 & 0.2184 & 0.2491 & 0.1807 & \textbf{0.1703}\\
DrivAerML (surface) & 60 & 0.1879 & 0.2212 & 0.1574 & \textbf{0.1513}\\
DrivAerML (surface) & 80 & 0.1709 & 0.2036 & 0.1417 & \textbf{0.1393}\\
DrivAerML (surface) & 100 & 0.1591 & 0.1842 & 0.1300 & \textbf{0.1275}\\
\midrule
DTCHull & 20 & 0.0707 & 0.0707 & 0.0632 & \textbf{0.0571}\\
DTCHull & 40 & 0.0601 & 0.0619 & 0.0568 & \textbf{0.0517}\\
DTCHull & 60 & 0.0577 & 0.0605 & 0.0527 & \textbf{0.0494}\\
DTCHull & 80 & 0.0555 & 0.0558 & 0.0485 & \textbf{0.0468}\\
DTCHull & 100 & 0.0539 & 0.0528 & 0.0476 & \textbf{0.0462}\\
\midrule
DrivAerML (volume) & 10 & 0.1174 & 0.1180 & 0.0926 & \textbf{0.0861}\\
DrivAerML (volume) & 20 & 0.0963 & 0.0998 & 0.0779 & \textbf{0.0743}\\
DrivAerML (volume) & 30 & 0.0776 & 0.0806 & 0.0662 & \textbf{0.0624}\\
DrivAerML (volume) & 40 & 0.0707 & 0.0739 & 0.0620 & \textbf{0.0590}\\
\bottomrule\end{tabular}\end{table}
\clearpage
\subsection{Data and Epoch Efficiency}
\label{app:efficiency}
\begin{figure}[!htb]\centering
\includegraphics[width=.88\linewidth]{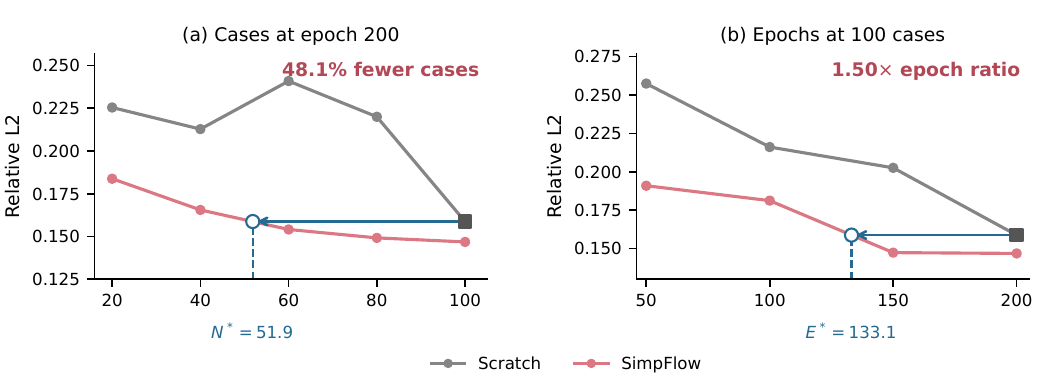}
\caption{Horizontal-intersection calculation on the AirCraft curves from \cref{fig:transfer}. A square marks scratch's final-budget, final-epoch error. The horizontal arrow meets SimpFlow at the same error (open circle); the vertical guide gives $N^*=51.9$ cases or $E^*=133.1$ epochs. These correspond to 48.1\% case saving and a $1.50\times$ epoch ratio.}
\label{fig:efficiencyinterpolation}
\end{figure}

The reference is scratch's error at the largest budget $N_{\max}$ and epoch 200. Moving horizontally left to each piecewise-linear curve gives its equivalent case budget $N^*$ (at epoch 200) or epoch $E^*$ (at $N_{\max}$), as illustrated in \cref{fig:efficiencyinterpolation}. We compute
\[
\text{Cases saved}=100\left(1-\frac{N^*}{N_{\max}}\right)\%,
\qquad \text{Epoch ratio}=\frac{200}{E^*}.
\]
We interpolate adjacent measured points on linear axes, taking the rightmost intersection if there are several. Fractional values are interpolated equivalents. A dash denotes an unreached target; a curve below target at its first point gives a bound without extrapolation. Scratch is the fixed endpoint reference. Epoch ratios exclude pre-training time.
\begin{table}[!ht]\centering\appendixtablestyle
\caption{Interpolated efficiency at the task-specific scratch reference. Case savings and epoch ratios use the same measured test grid as \cref{fig:transfer}.}
\label{tab:efficiency}
\begin{tabular}{llrrrr}\toprule
\textbf{Benchmark} & \textbf{Method} & \textbf{Equiv. cases} & \textbf{Cases saved (\%)} & \textbf{Equiv. epoch} & \textbf{Epoch ratio}\\\midrule
AirCraft & Scratch & 100.0 & 0.0 & 200.0 & 1.00\\
AirCraft & LGF & 86.9 & 13.1 & 145.7 & 1.37\\
AirCraft & GeoPT & 57.1 & 42.9 & 131.7 & 1.52\\
AirCraft & SimpFlow & 51.9 & 48.1 & 133.1 & 1.50\\
\midrule
NASA-CRM & Scratch & 100.0 & 0.0 & 200.0 & 1.00\\
NASA-CRM & LGF & -- & -- & -- & --\\
NASA-CRM & GeoPT & 76.1 & 23.9 & 148.2 & 1.35\\
NASA-CRM & SimpFlow & 64.7 & 35.3 & 136.9 & 1.46\\
\midrule
DrivAerML (surface) & Scratch & 100.0 & 0.0 & 200.0 & 1.00\\
DrivAerML (surface) & LGF & -- & -- & -- & --\\
DrivAerML (surface) & GeoPT & 58.6 & 41.4 & 131.8 & 1.52\\
DrivAerML (surface) & SimpFlow & 51.8 & 48.2 & 100.5 & 1.99\\
\midrule
DTCHull & Scratch & 100.0 & 0.0 & 200.0 & 1.00\\
DTCHull & LGF & 92.5 & 7.5 & 126.8 & 1.58\\
DTCHull & GeoPT & 54.0 & 46.0 & 100.7 & 1.99\\
DTCHull & SimpFlow & 31.7 & 68.3 & $\leq 50.0$ & $\geq 4.00$\\
\midrule
DrivAerML (volume) & Scratch & 40.0 & 0.0 & 200.0 & 1.00\\
DrivAerML (volume) & LGF & -- & -- & -- & --\\
DrivAerML (volume) & GeoPT & 26.1 & 34.7 & 106.4 & 1.88\\
DrivAerML (volume) & SimpFlow & 23.0 & 42.5 & 71.0 & 2.82\\
\bottomrule\end{tabular}\end{table}
\clearpage
\subsection{Output-Head Reset}
\label{app:headreset}
This control uses the main 4k initialization and the same query-validity mask, task splits, training schedules, and seed 42 as the main results, covering all 24 flow task--budget settings. We reset the final normalization and projection while retaining the input mapping and backbone, then fine-tune all parameters. At epoch 200, the reset model outperforms scratch and LGF in all 24 settings and GeoPT in 21. Its three losses to GeoPT occur on AirCraft at 80 and 100 cases and NASA-CRM at 100 cases, as with the inherited head. Thus, the main configuration's transfer gains do not depend on inheriting the output head.
\begin{table}[!ht]\centering\appendixtablestyle
\caption{Output-head control for the main 4k initialization, seed 42 and epoch 200. Entries are case-averaged test relative $L_2$. Both SimpFlow variants use the same pretrained checkpoint; only output-head initialization differs. Bold marks the lowest error in each row; ties at the displayed precision are both bold.}
\label{tab:headreset}
\begin{tabular}{llrrr}\toprule
\textbf{Task} & \textbf{Cases} & \textbf{GeoPT (reset)} & \textbf{SimpFlow 4k (inherited)} & \textbf{SimpFlow 4k (reset)}\\\midrule
AirCraft & 20 & 0.2063 & \textbf{0.1838} & 0.1875\\
AirCraft & 40 & 0.1685 & 0.1655 & \textbf{0.1628}\\
AirCraft & 60 & 0.1571 & 0.1541 & \textbf{0.1493}\\
AirCraft & 80 & \textbf{0.1489} & 0.1492 & 0.1511\\
AirCraft & 100 & \textbf{0.1335} & 0.1468 & 0.1453\\
\midrule
NASA-CRM & 20 & 0.2783 & \textbf{0.2213} & 0.2288\\
NASA-CRM & 40 & 0.1697 & 0.1592 & \textbf{0.1586}\\
NASA-CRM & 60 & 0.1622 & \textbf{0.1288} & 0.1316\\
NASA-CRM & 80 & 0.1169 & 0.1160 & \textbf{0.1154}\\
NASA-CRM & 100 & \textbf{0.1042} & 0.1054 & 0.1055\\
\midrule
DrivAerML surface & 20 & 0.2214 & 0.2128 & \textbf{0.2102}\\
DrivAerML surface & 40 & 0.1807 & \textbf{0.1703} & 0.1740\\
DrivAerML surface & 60 & 0.1574 & \textbf{0.1513} & 0.1517\\
DrivAerML surface & 80 & 0.1417 & \textbf{0.1393} & 0.1401\\
DrivAerML surface & 100 & 0.1300 & 0.1275 & \textbf{0.1260}\\
\midrule
DTCHull & 20 & 0.0632 & \textbf{0.0571} & 0.0586\\
DTCHull & 40 & 0.0568 & \textbf{0.0517} & 0.0528\\
DTCHull & 60 & 0.0527 & \textbf{0.0494} & 0.0504\\
DTCHull & 80 & 0.0485 & \textbf{0.0468} & 0.0474\\
DTCHull & 100 & 0.0476 & 0.0462 & \textbf{0.0457}\\
\midrule
DrivAerML volume & 10 & 0.0926 & \textbf{0.0861} & 0.0911\\
DrivAerML volume & 20 & 0.0779 & \textbf{0.0743} & 0.0767\\
DrivAerML volume & 30 & 0.0662 & \textbf{0.0624} & 0.0647\\
DrivAerML volume & 40 & 0.0620 & \textbf{0.0590} & 0.0596\\
\bottomrule\end{tabular}\end{table}

\clearpage
\subsection{Training and Selection Curves}
\label{app:learningcurves}
\begin{figure}[!ht]\centering
\includegraphics[width=\linewidth,height=.80\textheight,keepaspectratio]{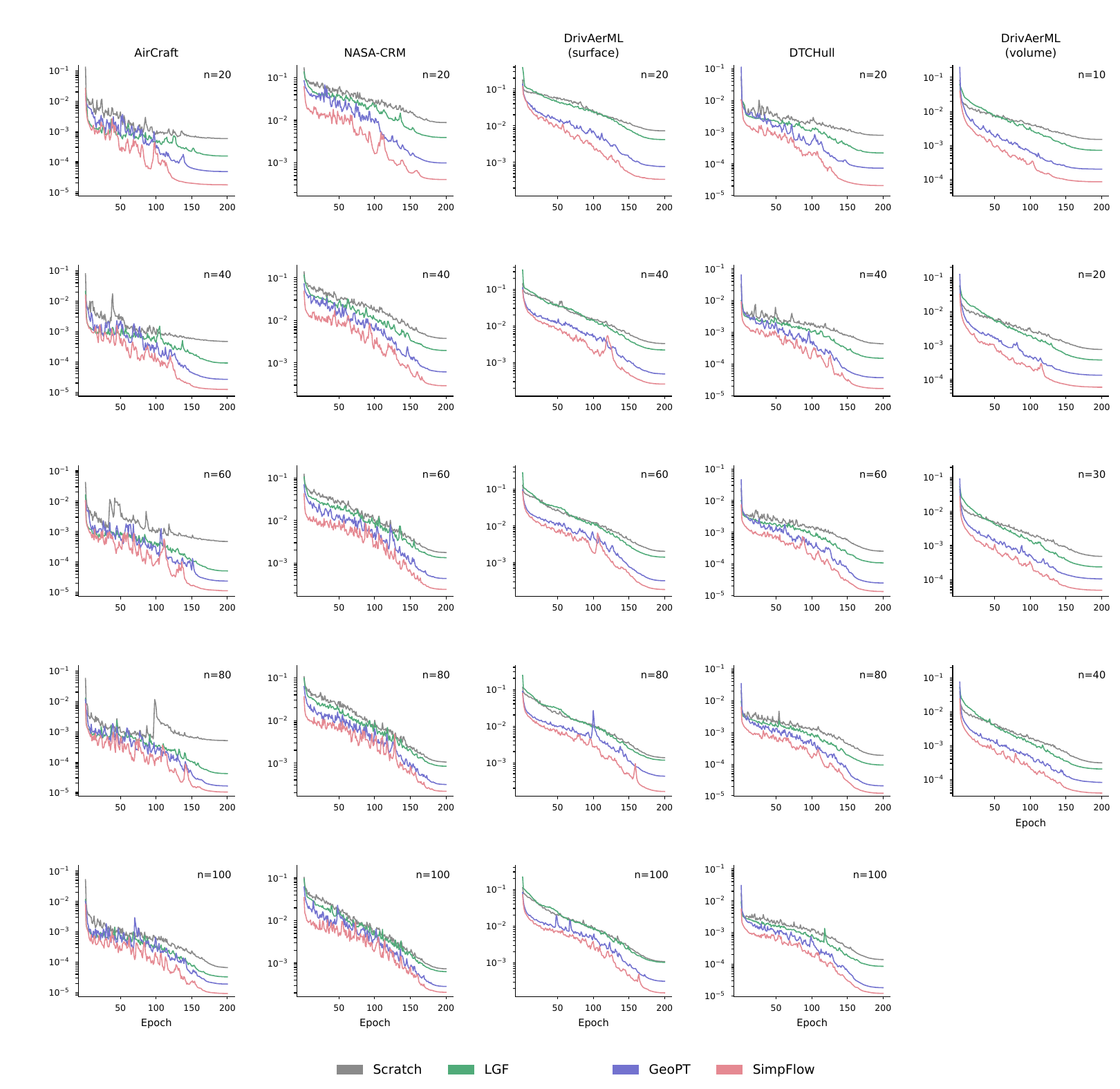}
\caption{Training MSE for seed 42. Columns are the five flow benchmarks; rows are their ordered budget levels. Empty cells indicate absent budgets. Vertical limits are shared across budgets within a task and use a logarithmic scale. Targets and normalizations differ across tasks. There is no across-seed uncertainty shading.}
\label{fig:traininghistory}\end{figure}\clearpage
\begin{figure}[!ht]\centering
\includegraphics[width=\linewidth,height=.80\textheight,keepaspectratio]{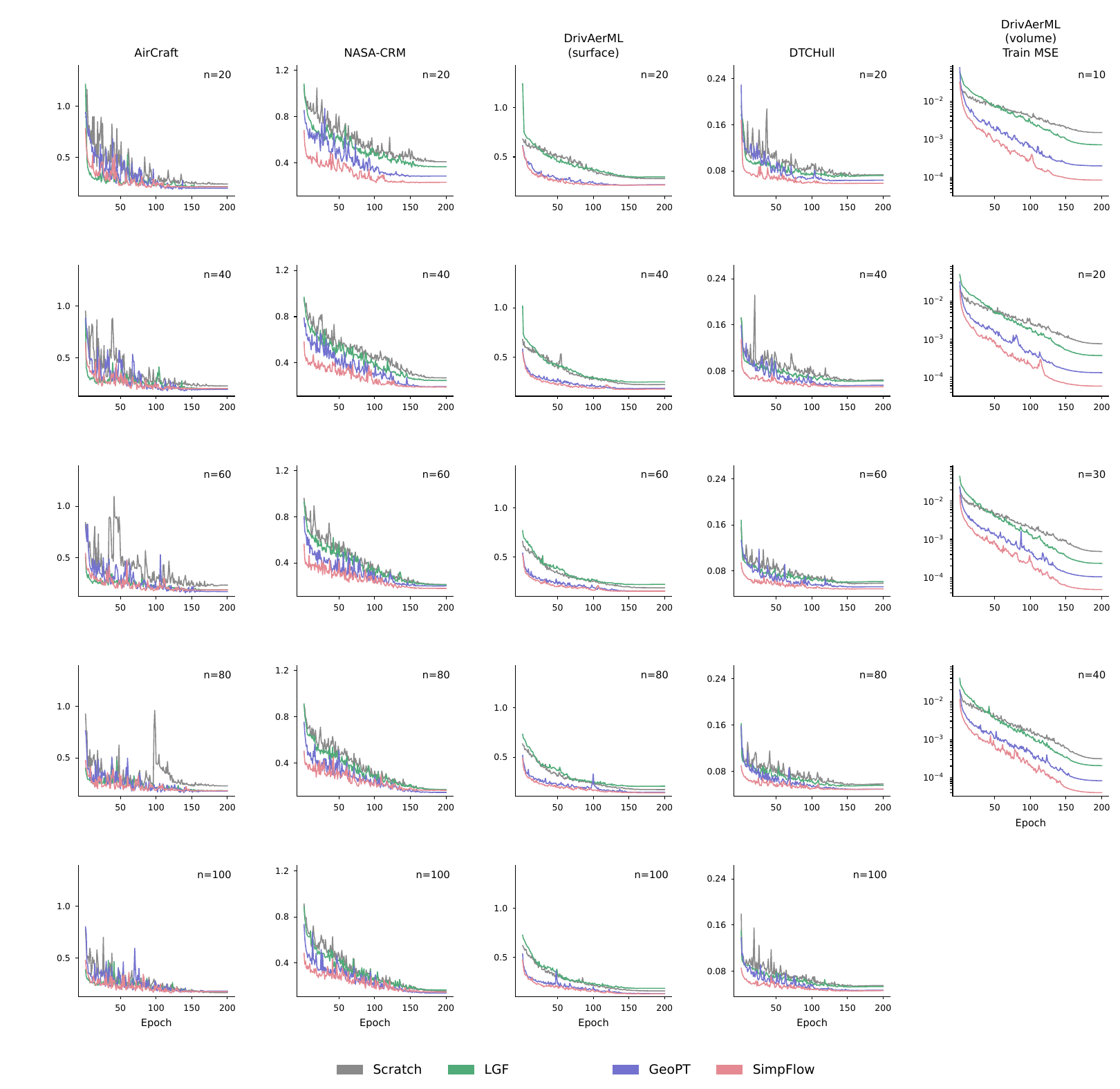}
\caption{Model-selection histories with the same layout as \cref{fig:traininghistory}. The four original tasks use validation relative $L_2$. DrivAerML volume has no validation set and uses training-set four-channel MSE, marked in its column title. These curves are separate from the fixed-checkpoint test grid and do not determine the reported epoch-200 results.}
\label{fig:validationhistory}\end{figure}

\clearpage
\subsection{Showcases}
\label{app:showcases}
We show the first test case of each benchmark at its largest training budget, comparing scratch, LGF, and SimpFlow (4k) (seed 42). These plots use the original model-selection checkpoints: training-selected for DrivAerML volume and validation-selected otherwise. They provide qualitative examples separately from the fixed epoch-200 comparisons in \cref{tab:fullresults}. Annotated relative $L_2$ values use all original supervised queries and match the saved evaluations.

Each pair of rows shows ground-truth and predicted fields followed by absolute errors. All methods share the same field color scale, with any display clipping stated in the caption; error scales saturate at the pooled 99th percentile. Surface and slice reconstruction use the same weights across methods, with a maximum support distance of 0.35 normalized units. Small car-surface gaps are interpolated from neighboring supported faces. Gray indicates unsupported surface regions or solid interiors; unsupported exterior regions are white.

\begin{figure}[!ht]
\centering
\includegraphics[width=\linewidth]{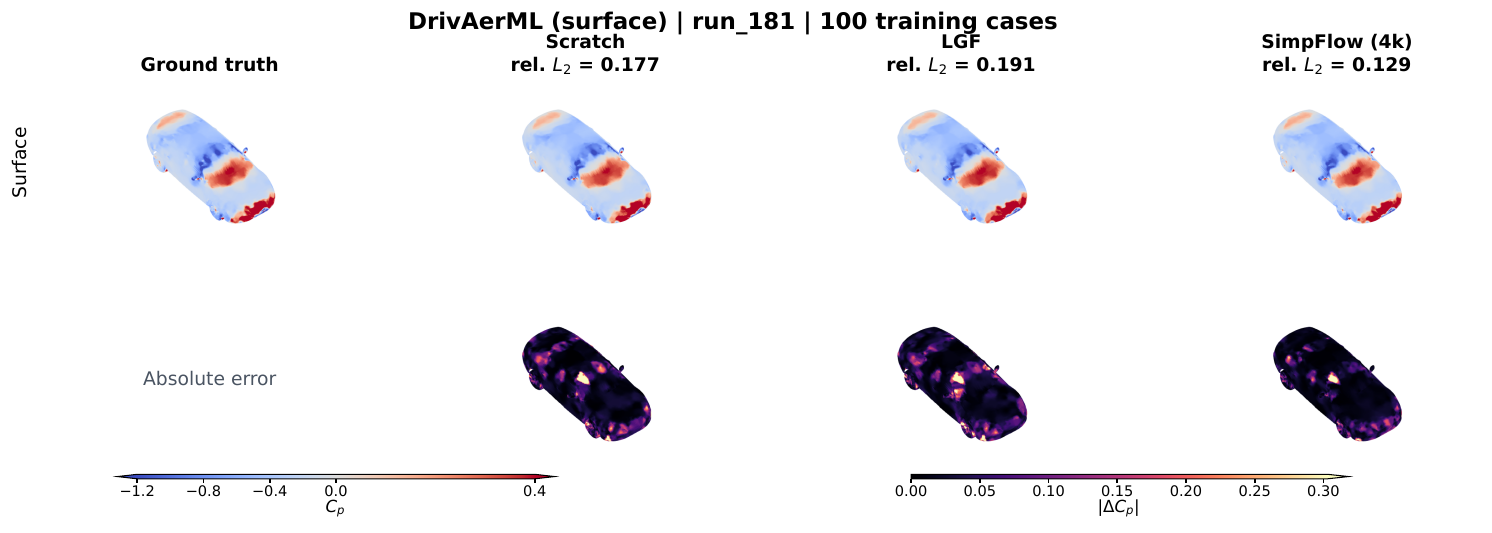}
\caption{DrivAerML surface pressure coefficient and absolute error for run\_181, with $U_\infty=38.889$\,m/s. Pressure colors saturate outside $[-1.2,0.4]$, with white at $C_p=0$. Relative $L_2$ uses all 4,096 surface queries.}
\label{fig:showcasecar}
\medskip
\includegraphics[width=\linewidth]{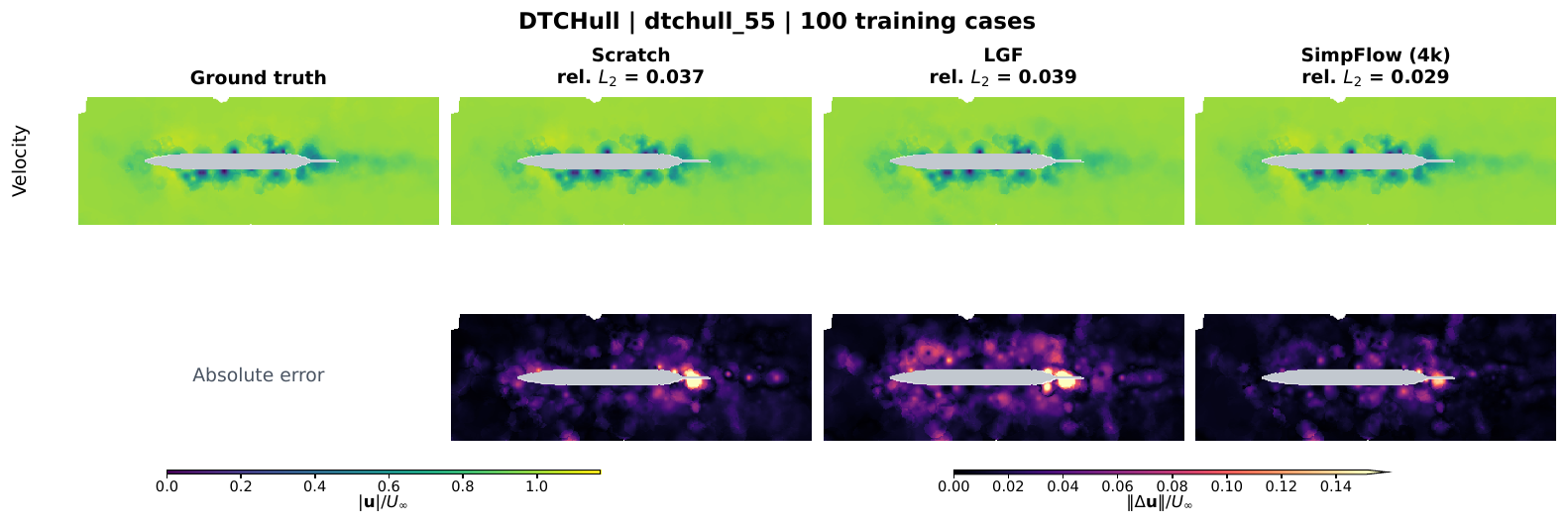}
\caption{DTCHull speed and velocity-vector error for dtchull\_55 at yaw $7.2075258^\circ$, on a horizontal slice 0.10 normalized units below the waterline. Relative $L_2$ uses the velocity vectors at all 3,584 supervised volume queries.}
\label{fig:showcasehull}
\end{figure}

\clearpage
\begin{figure}[!ht]
\centering
\includegraphics[width=.93\linewidth]{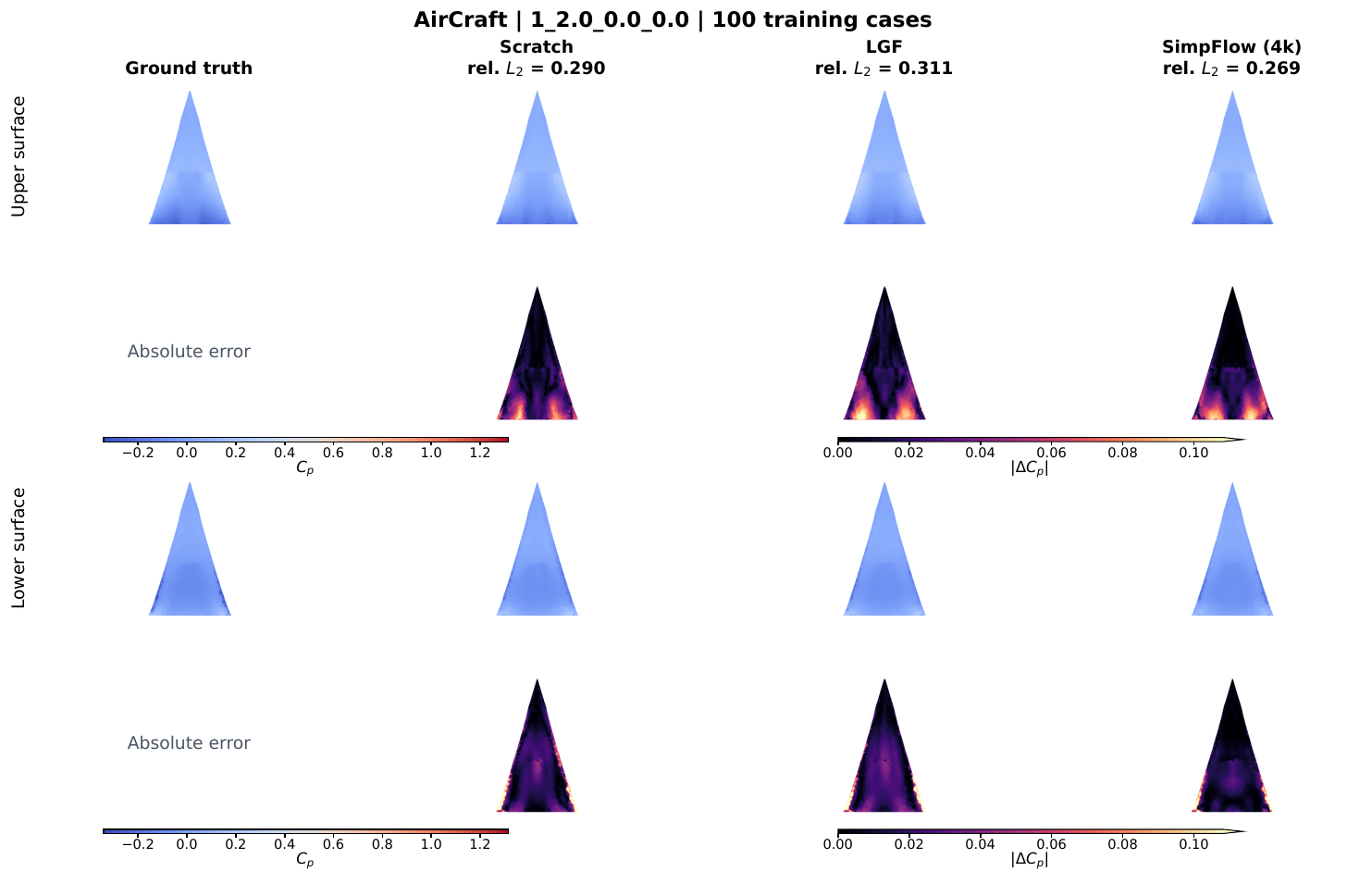}
\caption{AirCraft upper- and lower-surface pressure coefficients and errors for 1\_2.0\_0.0\_0.0, at Mach 2, zero angle of attack, and zero sideslip. Relative $L_2$ uses all 4,096 surface queries.}
\label{fig:showcaseaircraft}
\medskip
\includegraphics[width=.93\linewidth]{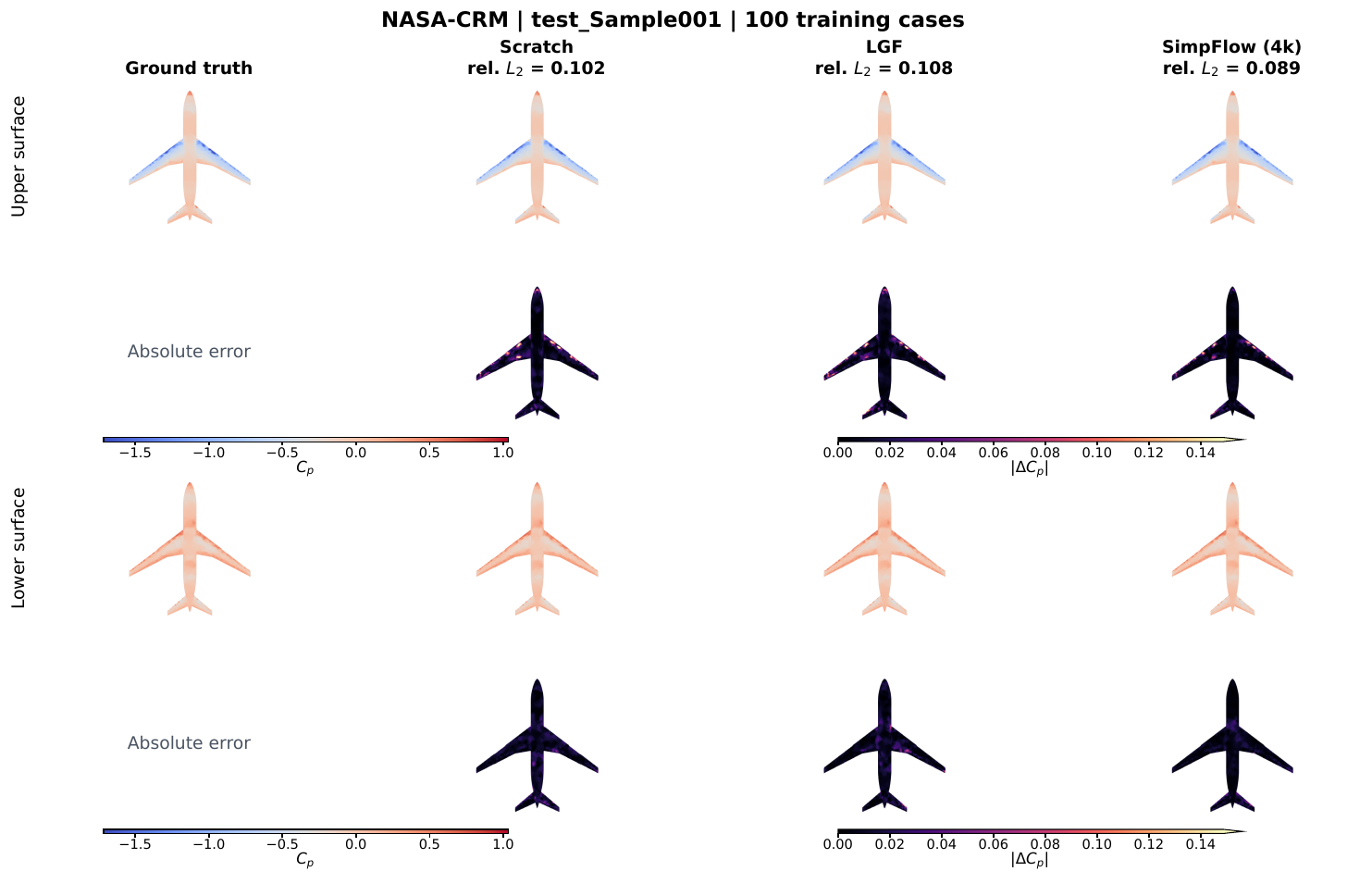}
\caption{NASA-CRM upper- and lower-surface pressure coefficients and errors for test\_Sample001, at Mach 0.721875 and angle of attack $3.45679^\circ$. Relative $L_2$ uses all 4,096 surface queries.}
\label{fig:showcasenasa}
\end{figure}

\clearpage
\begin{figure}[!ht]
\centering
\includegraphics[width=\linewidth]{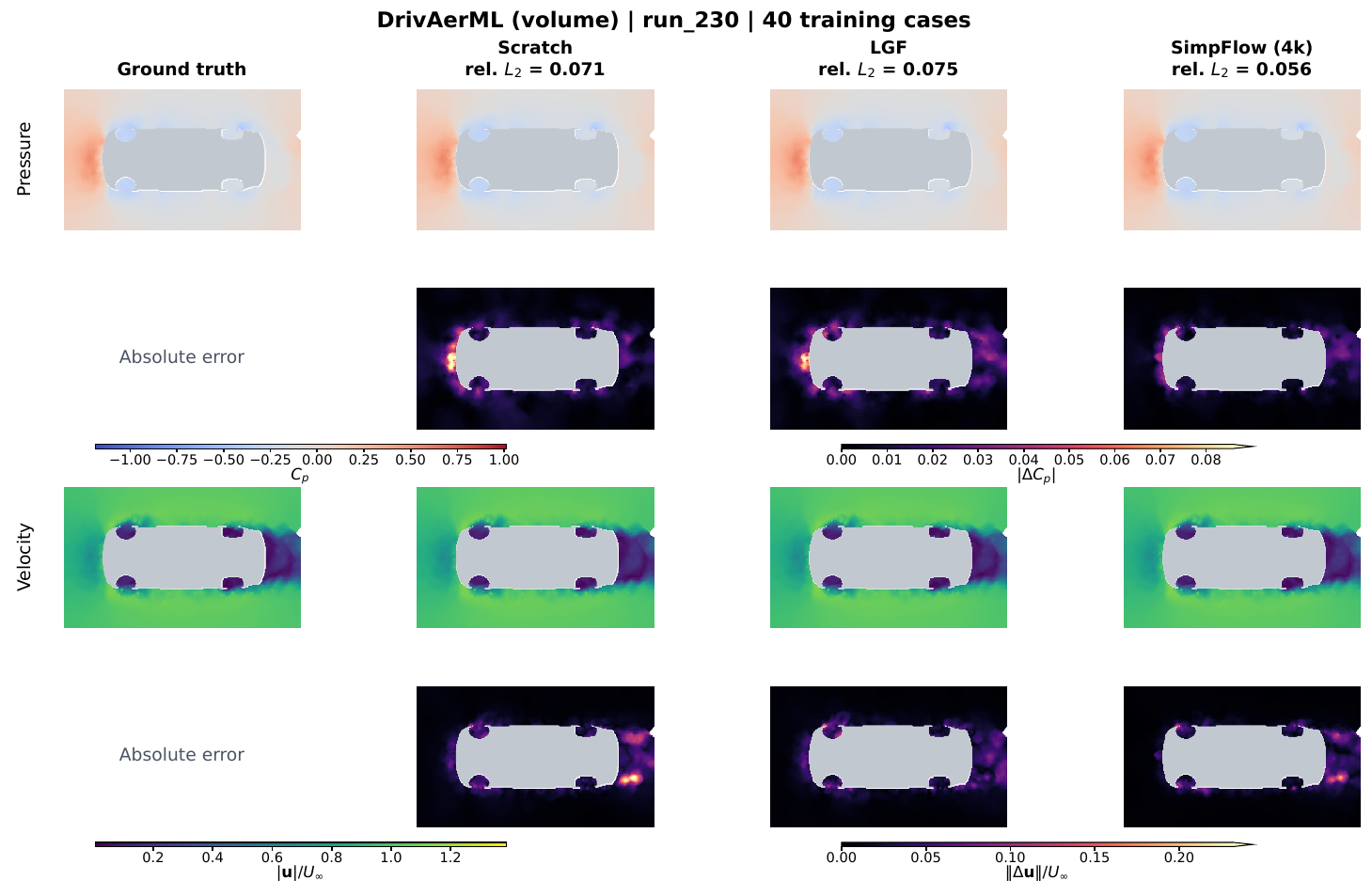}
\caption{DrivAerML volume pressure coefficient, speed, and their errors for run\_230, shown on a horizontal section through the car. Relative $L_2$ is the joint four-channel error over all 8,192 volume queries; the displayed velocity error is the norm of the vector difference.}
\label{fig:showcasevolume}
\medskip
\includegraphics[width=\linewidth]{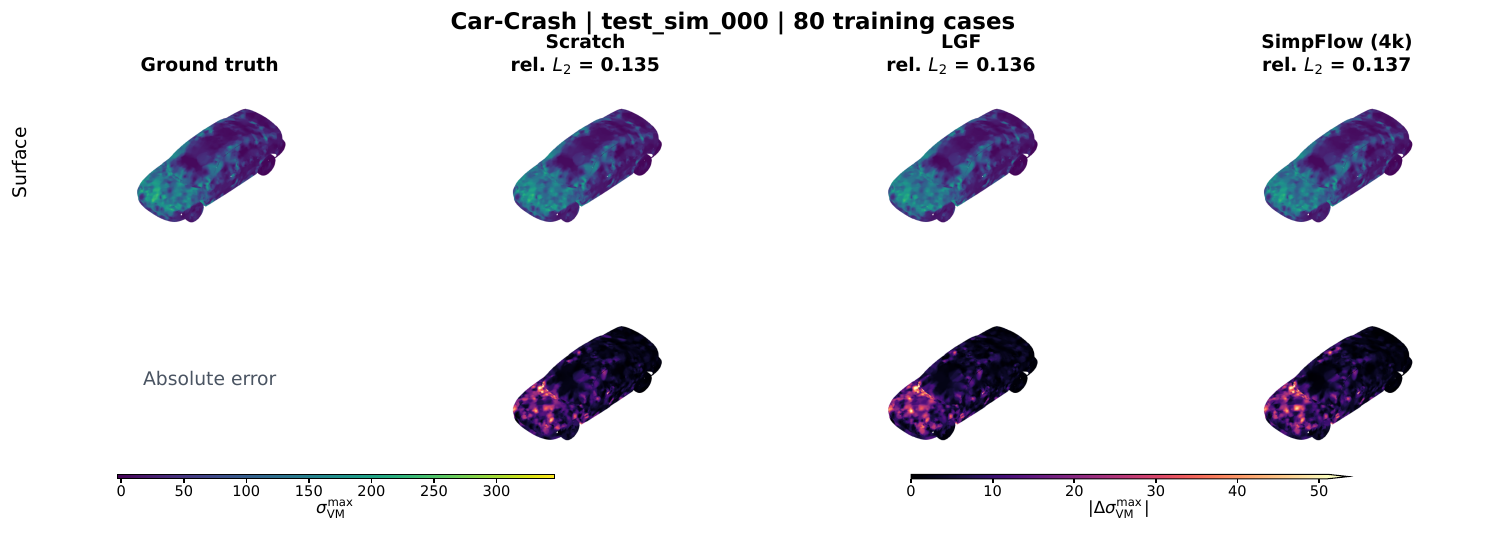}
\caption{Car-Crash maximum-over-time von Mises stress and absolute error for test\_sim\_000, displayed on the undeformed surface. Relative $L_2$ uses all 4,096 supervised surface queries, with stress restored to its original units.}
\label{fig:showcasecrash}
\end{figure}
\FloatBarrier

\FloatBarrier
\Needspace{.85\textheight}
\section{Transfer to Solid Mechanics}
\label{app:crashresults}
Car-Crash predicts maximum-over-time von Mises stress from the undeformed geometry. All pre-trained models reset their scalar stress head; SimpFlow uses the same 4k initialization as the flow comparisons. Protocols and normalization are in \cref{app:experiments}.

\subsection{Test Results}
\begin{table}[!ht]\centering\appendixtablestyle
\caption{Car-Crash test relative $L_2$ at epoch 200, seed 42. Bold marks the lowest error in each row.}
\label{tab:crashfinal}
\begin{tabular}{lrrrr}\toprule
\textbf{Cases} & \textbf{Scratch} & \textbf{LGF} & \textbf{GeoPT} & \textbf{SimpFlow}\\\midrule
20 & 0.5514 & 0.5389 & 0.5324 & \textbf{0.4786}\\
40 & \textbf{0.3151} & 0.3498 & 0.3297 & 0.3835\\
60 & \textbf{0.1719} & 0.1767 & 0.1775 & 0.1852\\
80 & 0.1242 & 0.1254 & \textbf{0.1231} & 0.1234\\
\bottomrule\end{tabular}\end{table}
At 20 training cases, SimpFlow lowers test error by 13.2\% relative to scratch. At 40 and 60 cases the error is 21.7\% and 7.7\% higher, respectively, while at 80 cases it is 0.7\% lower. The strongest benefit is therefore observed at the smallest training budget, with transfer benefits varying across budgets. For the scratch reference of 0.1242 at 80 cases and epoch 200, SimpFlow reaches the target at the same budget and epoch. \Cref{fig:crashtransfer} plots the budget and epoch curves.

\subsection{Pressure and Stress References}
The reference views below show the change from flow pressure to collision stress on the Neon geometry. The pressure illustration uses a 4k-panel solve, matching the resolution used to generate pretraining labels for the main initialization.
\begin{figure}[!ht]
\centering
\includegraphics[width=.70\textwidth]{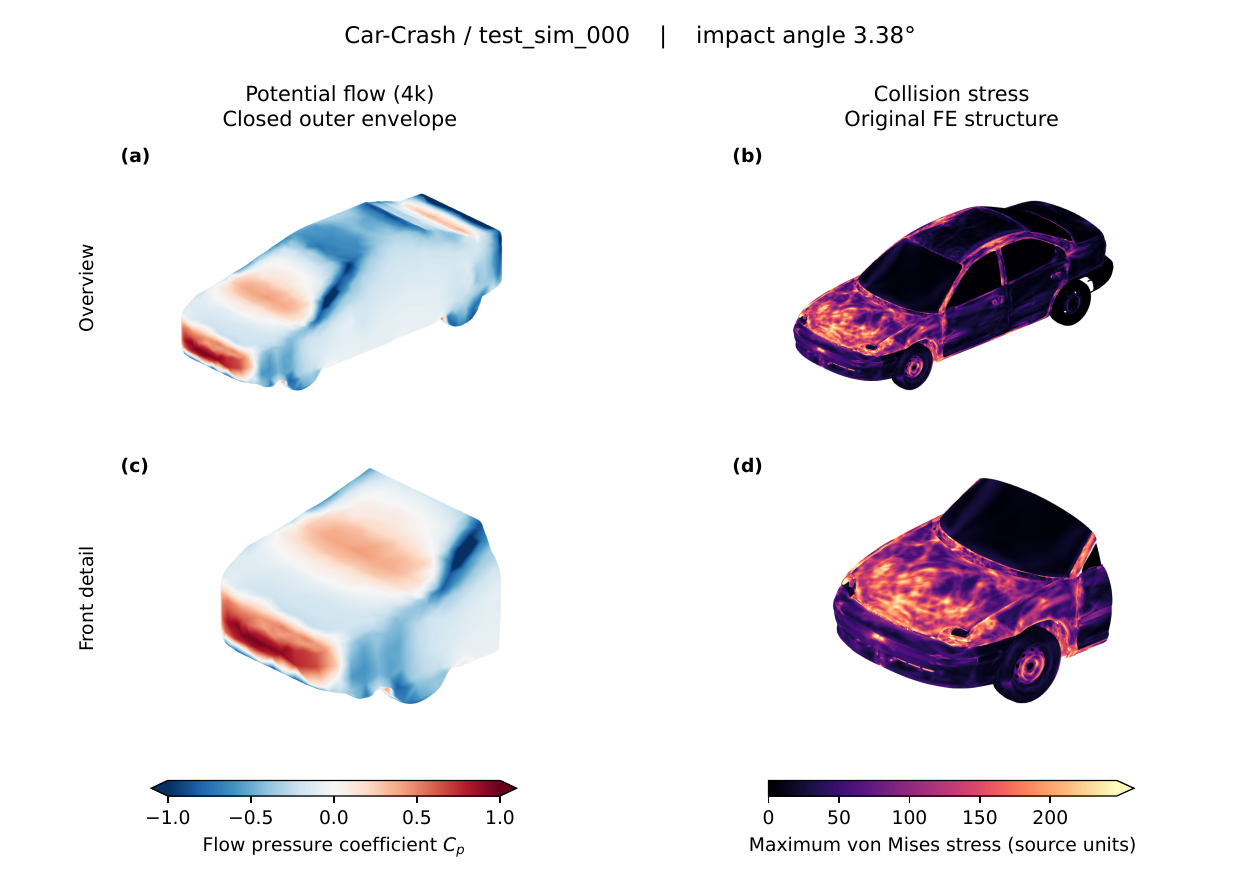}
\caption{Car-Crash \texttt{test\_sim\_000}, impact angle $3.38^\circ$, shown on the initial geometry. Left: illustrative 4k-panel potential-flow $C_p$ on a closed outer envelope constructed from geometry alone, with direction taken from the downstream condition encoding. Right: maximum-over-time von Mises stress on the original finite-element structure. Rows show an overview and front detail. Independent color limits are $[-1,1]$ for $C_p$ and $[0,247.05]$ in source-file units for stress, whose upper limit is the 99th percentile of original surface-cell values. For display, panel $C_p$ is averaged at vertices and interpolated within triangles onto the unchanged envelope. The envelope omits internal components. These are different physical quantities; no cross-quantity error is defined.}
\label{fig:crashreference}
\end{figure}
\FloatBarrier

\FloatBarrier
\Needspace{.75\textheight}
\section{Panel Resolution}
\label{app:ablations}
\subsection{Ablation Design}
\label{app:panelresolution}
We vary the target panel count across 500, 1k, 2k, 4k, and 8k to generate different numerical resolutions of the same potential-flow model. This ablation asks whether downstream transfer gains saturate as increasing the panel count raises the cost of generating pre-training labels. The dedicated 20-case 8k run shares the 2k/4k validity mask. It is separate from the original 8k initialization used in the broader budget comparison. The output-head reset control uses the main 4k initialization.

\subsection{Twenty-Case Transfer Results}
\begin{table}[!ht]
\centering\appendixtablestyle
\caption{Panel-resolution ablation. Entries are test relative $L_2$ at epoch 200, with 20 CFD training cases per benchmark and pre-training/fine-tuning seed 42. Counts denote target panel resolutions. Bold marks the lowest error in each row.}
\label{tab:panelresolution}
\begin{tabular}{lrrrrr}
\toprule
\textbf{Benchmark} & \textbf{500} & \textbf{1k} & \textbf{2k} & \textbf{4k} & \textbf{8k}\\
\midrule
AirCraft & 0.216012 & 0.192834 & 0.191728 & \textbf{0.183796} & 0.187804\\
NASA-CRM & 0.286118 & 0.253863 & 0.221480 & \textbf{0.221267} & 0.236149\\
DrivAerML & 0.277265 & 0.254397 & 0.217776 & \textbf{0.212799} & 0.219389\\
DTCHull & 0.062610 & 0.058658 & 0.056837 & 0.057138 & \textbf{0.056514}\\
\bottomrule
\end{tabular}
\end{table}

The comparison in \cref{fig:panelresolution} reports the relative reduction in downstream test error,
\begin{equation}
G_b(M)=100\,\frac{E_b^{\mathrm{scratch}}-E_b^{(M)}}{E_b^{\mathrm{scratch}}},
\end{equation}
where $E_b^{(M)}$ is the epoch-200 case-averaged test relative $L_2$. Scratch uses the same subset, seed, and evaluation cases, with errors 0.225441, 0.391862, 0.286293, and 0.070693 in table order. All five resolutions improve over scratch. The 500-panel error exceeds 1k on every task, while 2k--8k shows no monotonic improvement.

\subsection{Transfer Across Budgets}
\Cref{tab:resolutionwins} compares three available initializations on the same 22 task--budget settings: five budgets per original flow task and 20/40 cases for volume DrivAerML. Strictly lower epoch-200 test error counts as a win: 15/22 for 2k, 19/22 for 4k, and 19/22 for the original 8k model. The additional 10/30-case volume runs appear only in the 4k main grid and are outside this common set. This broader comparison uses the original 8k mask, while the dedicated 20-case 8k control above shares the 2k/4k mask. Labeling costs for these resolutions follow the common timing protocol in \cref{app:cost}.

\end{document}